\documentclass[12pt,aps,prb,showpacs,amsmath,amssymb,superscriptaddress,floatfix]{revtex4-2}

\usepackage[english]{babel}
\usepackage[utf8]{inputenc}
\usepackage{graphicx}
\usepackage{bm} 
\usepackage{latexsym}
\usepackage{amsfonts}
\usepackage{bbold}
\usepackage{appendix}
\usepackage[colorlinks = true,
            linkcolor = blue,
            urlcolor  = blue,
            citecolor = blue,
            anchorcolor = blue]{hyperref}

\newcommand{\sg}{\sigma}

\newcommand{\Gam}{\Gamma}
\newcommand{\lam}{\lambda}
\newcommand{\Lam}{\Lambda}

\newcommand{\up}{\uparrow}
\newcommand{\down}{\downarrow}

\newcommand{\bk}{\mathbf{k}}
\newcommand{\bq}{\mathbf{q}}
\newcommand{\bQ}{\mathbf{Q}}
\newcommand{\br}{\mathbf{r}}

\newcommand{\bS}{\mathbf{S}}

\newcommand{\cA}{\mathcal{A}}

\newcommand{\cD}{\mathcal{D}}

\newcommand{\cJ}{\mathcal{J}}

\newcommand{\cP}{\mathcal{P}}
\newcommand{\cR}{\mathcal{R}}
\newcommand{\cS}{\mathcal{S}}

\newcommand{\tr}{{\rm tr}}
\newcommand{\lra}{\leftrightarrow}
\newcommand{\bra}{\langle}
\newcommand{\ket}{\rangle}

\begin{document}

\author{Paulo Forni}
\affiliation{Max Planck Institute for Solid State Research, D-70569 Stuttgart, Germany}

\author{Pietro M.~Bonetti}
\affiliation{Department of Physics, Harvard University, Cambridge MA 02138, USA}

\author{Henrik M\"uller-Groeling}
\affiliation{Max Planck Institute for Solid State Research, D-70569 Stuttgart, Germany}

\author{Demetrio Vilardi}
\affiliation{Max Planck Institute for Solid State Research, D-70569 Stuttgart, Germany}

\author{Walter Metzner}
\affiliation{Max Planck Institute for Solid State Research, D-70569 Stuttgart, Germany}

\title{Spin susceptibility in a pseudogap state with fluctuating spiral magnetic order}
\date{\today}
\begin{abstract}
We compute the electron spin susceptibility in the pseudogap regime of the two-dimensional Hubbard model in the framework of a SU(2) gauge theory of fluctuating magnetic order.
The electrons are fractionalized in fermionic chargons with a pseudospin degree of freedom and bosonic spinons. The chargons are treated in a renormalized mean-field theory and order in a N\'eel or spiral magnetic state in a broad range around half-filling below a transition temperature $T^*$. Fluctuations of the spin orientation are captured by the spinons. Their dynamics is governed by a non-linear sigma model, with spin stiffnesses computed microscopically from the pseudospin susceptibility of the chargons.
The SU(2) gauge group is higgsed in the chargon sector, and the spinon fluctuations prevent breaking of the physical spin symmetry at any finite temperature.
The electron spin susceptibility obtained from the gauge theory shares many features with experimental observations in the pseudogap regime of cuprate superconductors: the dynamical spin susceptibility $S(\bq,\omega)$ has a spin gap, the static uniform spin susceptibility $\kappa_s$ decreases strongly with temperature below $T^*$, and the NMR relaxation rate $T_1^{-1}$ vanishes exponentially in the low temperature limit if the ground state is quantum disordered.
At low hole doping, $S(\bq,\omega)$ exhibits nematicity below a transition temperature $T_{\rm nem} < T^*$, and at larger hole doping in the entire pseudogap regime below $T^*$.
\end{abstract}
\maketitle
%


\section{Introduction}

A peculiar feature of cuprate high temperature superconductors is their pseudogap behavior above the critical temperature for superconductivity, observed in a broad doping range from the underdoped into the optimally doped regime \cite{Keimer2015, Proust2019}.
It is characterized by a reduction of the charge carrier concentration, a spin gap, and a reconstructed Fermi surface which in photoemission looks like a collection of Fermi arcs. It is often accompanied by electronic nematicy, breaking the tetragonal symmetry of the crystal. 

There is convincing numerical evidence for pseudogap behavior in the two-dimensional Hubbard model, in particular from quantum cluster calculations \cite{Qin2022}, and fluctuation diagnostics of the contributions to the self-energy has revealed that the pseudogap is generated mainly by antiferromagnetic spin fluctuations \cite{Gunnarsson2015}.
While the guidance provided by such numerical simulations is very valuable, a deeper understanding of the pseudogap phenomenon and data with a higher momentum resolution remain desirable.

Approximate analytic theories can provide further insights on the structure of the pseudogap state, and information on its fine-structure in momentum space. Early theories of the pseudogap phenomenon were based on weak-coupling expansions, such as Moriya's theory \cite{Moriya2000} and the two-particle self-consistent approach by Vilk and Tremblay \cite{Vilk1996}. The Mermin-Wagner theorem on the absence of spin symmetry breaking at finite temperatures is respected in these theories, but the pseudogap forms only for large magnetic correlation lengths and thus in a limited parameter regime, while the numerical results show that strong short-ranged correlations are actually sufficient.
Noteworthy previous theories involving only short-ranged correlations are based on Anderson's resonanting valence bond (RVB) idea \cite{Lee2006}, and on dynamical gap generation by umklapp processes, as proposed by Yang, Rice, and Zhang \cite{Yang2006}.

More recently it was shown that the pseudogap behavior observed in cuprates can be captured by SU(2) gauge theories of fluctuating magnetic order \cite{Sachdev2016, Chatterjee2017, Scheurer2018, Wu2018, Sachdev2019, Bonetti2022a}. Formally this approach is based on a fractionalization of the electron into a fermionic {\em chargon}\/ with a pseudospin degree of freedom and a charge neutral {\em spinon}, that is, a SU(2) matrix providing a space and time dependent local reference frame \cite{Schulz1995}. The local spin rotations can be parametrized by a SU(2) gauge field, and the fractionalization leads to a gauge redundancy. One can then construct states where the chargons exhibit some sort of magnetic order (N\'eel, spiral, or stripe), while the spinon fluctuations restore the SU(2) spin symmetry and prevent magnetic long-range order of the physical spin-carrying electrons \cite{Dupuis2002, Borejsza2004, Sachdev2009}, at least at finite temperatures. Quantities involving only charge degrees of freedom behave essentially as in a conventional magnetically ordered state, and the Fermi surface gets correspondingly reconstructed.
Recently a more sophisticated gauge theory of the pseudogap regime with additional auxiliary degrees of freedom (``ancilla qubits'') was proposed \cite{Zhang2020, Zhang2020a, Mascot2022, Nikolaenko2023, Bonetti2024}.

In this paper we analyze the dynamical spin susceptibility as obtained from the SU(2) gauge theory of fluctuating magnetic order in a pseudogap state. Our analysis builds on the formalism of Ref.~\cite{Bonetti2022a}, which is designed to quantitatively compute physical observables for a microscopic model such as the two-dimensional Hubbard model. We focus on regions in the phase diagram where the chargons order in a circular spiral or in a N\'eel state. Pseudogap behavior emerging from stripe states will be analyzed in subsequent work. We present results for the dynamical spin susceptibility $S(\bq,\omega)$ and its static uniform limit, and for the NMR relaxation rate.

The remainder of the article is structured as follows. In Sec.~II we review the SU(2) gauge theory of fluctuating spiral/N\'eel order. General expressions for the electron spin susceptibility are derived in Sec.~III, while concrete results are presented and discussed in Sec.~IV. A conclusion with a summary and outlook in Sec.~V closes the presentation.


\section{SU(2) gauge theory}
\label{sec: gauge theory}

We consider the Hubbard model on a square lattice \cite{Arovas2022, Qin2022} for the sake of concreteness. Extensions to extended Hubbard models (with non-local interactions) and other lattice geometries are straightforward.
Our analysis is based on the functional integral formalism \cite{Negele1987}.
We use natural units so that $\hbar = k_B = 1$.
The Hubbard action has the form
\begin{eqnarray} \label{eq: Hubbard action}
 \mathcal{S}[c,c^*] &=&
 \int_0^\beta\!d\tau \sum_{j,j',\sigma} c_{j\sg}^*(\tau)
 \left[ (\partial_\tau - \mu) \delta_{jj'} + t_{jj'} \right] c_{j'\sg}(\tau)
 \nonumber \\
 &+& \int_0^\beta\!d\tau \; U \sum_j n_{j\up}(\tau) \, n_{j\down}(\tau) ,
\end{eqnarray}
where $c_{j\sg}(\tau)$ and $c_{j\sg}^*(\tau)$ are imaginary time Grassmann fields corresponding to the annihilation and creation, respectively, of an electron with spin orientation $\sg$ at site $j$, and
$n_{j\sg}(\tau) = c_{j\sg}^*(\tau) c_{j\sg}(\tau)$.
The hopping amplitudes $t_{jj'}$ are translation invariant.
The chemical potential is denoted by $\mu$, and $U > 0$ is the strength of the Hubbard interaction.
The Hubbard action is invariant under \emph{global} SU(2) spin rotations
$c_j(\tau) \to \mathcal{U} c_j(\tau)$,
$c_j^*(\tau) \to c_j^*(\tau) \, \mathcal{U}^\dag$,
where $c_j(\tau)$ and $c^*_j(\tau)$ are two-component spinors composed from $c_{j\sg}(\tau)$ and $c^*_{j\sg}(\tau)$ with $\sg \in \{\up,\down\}$, respectively, while $\mathcal{U}$ is an arbitrary SU(2) matrix.


\subsection{Electron fractionalization}

To separate spin orientation fluctuations from the charge and spin amplitude degrees of freedom, we fractionalize the electron fields as~\cite{Schulz1995, Dupuis2002, Borejsza2004, Sachdev2009}
\begin{equation} \label{eq: electron fract}
 c_j(\tau) = R_j(\tau) \, \psi_j(\tau)  \, , \quad
 c_j^*(\tau) = \psi_j^*(\tau) \, R_j^\dag(\tau) ,
\end{equation}
where $R_j(\tau) \in \mbox{SU(2)}$, to which we refer as {\em spinon}\/, is composed of bosonic fields, and the components $\psi_{js}$ of the {\em chargon}\/ spinor $\psi_j$ are fermionic.
The factorization in Eq.~\eqref{eq: electron fract} introduces a redundant SU(2) symmetry associated with the pseudospin of the chargons.

Using Eq.~\eqref{eq: electron fract}, the Hubbard action can be expressed in terms of the spinon and chargon fields. The quadratic part of Eq.~\eqref{eq: Hubbard action} can be written as \cite{Borejsza2004}
\begin{eqnarray} \label{eq: S0 chargons spinons}
 \mathcal{S}_0[\psi,\psi^*,R] &=& \int_0^\beta\!d\tau
 \bigg\{ \sum_j \psi^*_j(\tau) \left[ \partial_\tau - \mu - \phi_j(\tau) \right]
 \psi_{j}(\tau) \nonumber \\
 && + \, \sum_{j,j'} t_{jj'} \, \psi_j^*(\tau) \, e^{iA_{jj'}(\tau)} \,
 \psi_{j'}(\tau) \bigg\},
\end{eqnarray}
with an SU(2) gauge field $(\phi,A)$ related to the spinon rotations by
$\phi_j(\tau) = - R_j^\dag(\tau) \partial_\tau R_j(\tau)$ and
$e^{iA_{jj'}(\tau)} = R_j^\dag(\tau) R_{j'}(\tau)$.
Both $\phi_j(\tau)$ and $A_{jj'}(\tau)$ are elements of the Lie algebra of SU(2), and can therefore be written as linear combinations of the Pauli matrices $\sg^a$ with $a = 1,2,3$:
$\phi_j(\tau) = \frac{1}{2} \sum_a \phi_j^a(\tau) \sigma^a$ and
$A_{jj'}(\tau) = \frac{1}{2} \sum_a A_{jj'}^a(\tau) \sigma^a$.
In the interaction part of the Hubbard action \eqref{eq: Hubbard action}, the SU(2) rotations drop out \cite{Bonetti2022a}, so that
\begin{equation} \label{eq: S_int}
 \mathcal{S}_I[\psi,\psi^*,R] = \mathcal{S}_I[\psi,\psi^*] =
 U \int_0^\beta\!d\tau \sum_j n^\psi_{j\up}(\tau) \, n^\psi_{j\down}(\tau),
\end{equation}
with $n^\psi_{js}(\tau) = \psi^*_{js}(\tau) \, \psi_{js}(\tau)$.
The action $\mathcal{S} = \mathcal{S}_0 + \mathcal{S}_I$ represents the Hubbard model action where the physical electrons have been replaced by chargons coupled to an SU(2) gauge field.


\subsection{Spinon effective action}

Our strategy is to treat the chargons in an approximation that ignores fluctuations of the spin orientation, and treat the latter by the spinon degrees of freedom.
Assigning only long wavelength fluctuations to the rotation matrices $R_j(\tau)$, we can assume that they vary slowly in space. Under this assumption a continuum limit and a gradient expansion can be applied. The latter corresponds to an expansion in powers of the gauge fields.

Expanding to second order in the gauge fields, and integrating out the chargon fields, one can derive an effective action for the spinons of the form \cite{Dupuis2002, Bonetti2022a}
\begin{equation} \label{eq: eff action}
 \cS[A] = \frac{1}{2} \sum_{a,b} J_{\mu\nu}^{ab} \int_q A_\mu^a(q) \, A_\nu^b(-q) \, ,
\end{equation}
where $A_\mu^a(q)$ with $\mu \in \{0,x,y\}$ and $q = (q_0,q_x,q_y)$ is the gauge field in momentum representation.
We use Einstein's summation convention for repeated space-time indices $\mu$ and $\nu$, but not for the SU(2) indices $a$ and $b$. The integral is written in a shorthand notation
$\int_q = T \sum_{q_0} \int \frac{d^2\bq}{(2\pi)^2}$, where $q_0$ is a bosonic Matsubara frequency.
The coefficients $J_{\mu\nu}^{ab}$ in Eq.~\eqref{eq: eff action} are the \emph{spin stiffnesses} associated with the Goldstone modes of the magnetically ordered chargon state.
They can be computed either from the spin susceptibility of the chargon state, or from its SU(2) gauge field response kernel \cite{Bonetti2022a, Bonetti2022, Bonetti2022ward, Goremykin2024}.

Fourier transforming to position space $x=(\br,\tau)$, the action \eqref{eq: eff action} can be mapped to a non-linear sigma model of the form \cite{Bonetti2022a}
\begin{equation} \label{eq: sigma model}
 \cS[\cR] = \frac{1}{2} \int dx \, \tr \left\{ \cP_{\mu\nu}
 [\partial_\mu \cR^T(x)] [\partial_\nu \cR(x)] \right\} \, ,
\end{equation}
where $\cR$ is the \emph{adjoint} representation of the SU(2) rotation $R$, defined by
\begin{equation} \label{eq: adjoint rep}
 R^\dag \sg^a R = \sum_b \cR^{ab} \sg^b \, ,
\end{equation}
and $\cP_{\mu\nu} = \frac{1}{2} \tr(\cJ_{\mu\nu}) \, I_3 - \cJ_{\mu\nu}$, with the $3\times3$ identity matrix $I_3$, and the $3\times3$ matrix $\cJ_{\mu\nu}$ formed by the matrix elements $J_{\mu\nu}^{ab}$.


\subsection{Chargons} \label{sec: chargons}


\subsubsection{Renormalized mean-field theory}

We treat the chargons in a renormalized mean-field theory \cite{Wang2014, Yamase2016}. Mean-field equations for ordered phases are thereby solved with renormalized instead of bare interactions as input parameters. The renormalized interactions are obtained from a functional renormalization group flow \cite{Metzner2012}, which takes charge, spin, and pairing channels into account on equal footing. The coupled fluctuations of the various channels have two important effects: i) magnetic and charge interactions are reduced, and ii) d-wave pairing interactions are generated by fluctuations \cite{Metzner2012}. Since we are interested in the normal (non-superconducting) state, we ignore the pairing instability by assuming that the temperature is above the transition temperature for pairing, or that superconductivity has been suppressed by an external magnetic field.

The effective interactions are momentum and frequency dependent, but these dependences are rather weak for the {\em two-particle irreducible}\/ \cite{Wang2014} effective magnetic interaction which enter the mean-field equation for the magnetic order parameter. Hence we compute the effective interaction in the magnetic channel at zero frequency and at the wave vector where the magnetic interaction is maximal, and use this as a momentum and frequency independent coupling constant $\bar U$.


\subsubsection{Spiral magnetic order}
\label{sec: spiral order}

Numerous static and dynamical mean-field calculations for the two-dimensional Hubbard model yield planar spiral magnetic order in a sizable portion of the phase diagram \cite{Fresard1991, Chubukov1992, Chubukov1995, Igoshev2010, Bonetti2020, Goremykin2023}.
Choosing the spins to be aligned in the $xy$ plane (all planes in spin space are energetically degenerate), the spin expectation value in a spiral state has the time-independent form
\begin{equation} \label{eq: spiral}
 \langle \bS_j^\psi(\tau) \rangle = m \left[
 \cos(\bQ \cdot \br_j) \, {\bf e}_x + \sin(\bQ \cdot \br_j) \, {\bf e}_y
 \right] \, ,
\end{equation}
where $\bQ$ is a fixed wave vector, $m$ is the spin amplitude, $\br_j$ is the position vector of the lattice site $j$, and ${\bf e}_\alpha$ are unit vectors with directions $\alpha = x,y,z$.
The components of the pseudospin field of the chargons $\bS_j^\psi(\tau)$ can be expressed in terms of the chargon fields as
\begin{equation} \label{eq: Sj}
 S^{\psi a}_j(\tau) = \frac{1}{2} \sum_{s,s'}
 \psi_{js}^*(\tau) \sigma_{ss'}^a \psi_{js'}(\tau) \, ,
\end{equation}
where $\sigma^a$ are the Pauli matrices.

Diagonalizing the (renormalized) mean-field Hamiltonian for the spiral state yields two quasiparticle bands of the form
\begin{equation}
 E_\bk^\pm = \frac{1}{2} (\xi_\bk + \xi_{\bk+\bQ}) \pm
 \sqrt{\frac{1}{4}(\xi_\bk - \xi_{\bk+\bQ})^2 + \Delta_m^2} \, ,
\end{equation}
where $\Delta_m$ is the energy gap associated with the spiral magnetic order.
Extracting an effective (renormalized) interaction $\bar U$ from the functional renormalization group flow \cite{Wang2014, Yamase2016}, the gap is determined by the self-consistency equation
\begin{equation}
 \Delta_m = {\bar U} \, m = {\bar U} \, \Delta_m \int_\bk
 \frac{f(E_\bk^-) - f(E_\bk^+)}{E_\bk^+ - E_\bk^-} \, ,
\end{equation}
where $f(x) = (e^{x/T} + 1)^{-1}$ is the Fermi function.
The optimal wave vector $\bQ$ is determined by minimizing the free energy
\begin{equation}
 F(\bQ) = -T \int_\bk \sum_{\ell=\pm} \ln \big[ 1 + e^{-E_\bk^\ell(\bQ)/T} \big]
 + {\bar U}^{-1} \Delta_m^2(\bQ) + \mu(\bQ) n \, ,
\end{equation}
where the chemical potential $\mu(\bQ)$ is determined by keeping the electron density $n$ fixed.
For our choice of model parameters, the optimal $\bQ$ always has one component equal to $\pi$, that is, the wave vector has the form $\bQ = (\pi - 2\pi\eta,\pi)$, or symmetry related, with $\eta \geq 0$.

It is convenient to introduce rotated fermion fields \cite{Kampf1996}
\begin{equation} \label{eq: spinrot}
\begin{split}
 \tilde{\psi}_j(\tau) &=
 e^{-\frac{i}{2} \bQ \cdot \br_j} e^{\frac{i}{2} \bQ \cdot \br_j \sigma^3}
 \psi_j(\tau) \, , \\
 \tilde{\psi}_j^*(\tau) &= \psi_j^*(\tau) \,
 e^{-\frac{i}{2} \bQ \cdot \br_j \sigma^3} e^{\frac{i}{2} \bQ \cdot \br_j } \, .
\end{split}
\end{equation}
In this rotated basis, the spins appear aligned ferromagnetically in the $x$ direction, so that the system remains translation invariant.
In momentum space, the transformation \eqref{eq: spinrot} corresponds to
$\tilde{\psi}_\bk = (\tilde{\psi}_{\bk\up},\tilde{\psi}_{\bk\down}) =
(\psi_{\bk\up},\psi_{\bk+\bQ\down})$.
The rotated chargon Green function is diagonal in momentum space (corresponding to translation invariance in real space), and has the form
\begin{equation}
 \widetilde{G}^{-1}(\bk,ik_0) = \left( \begin{matrix}
 ik_0 - \xi_\bk & - \Delta_m \\ - \Delta_m & ik_0  - \xi_{\bk+\bQ}
 \end{matrix} \right) \, ,
\end{equation}
where $k_0$ denotes a fermionic Matsubara frequency.
Using a partial fraction decomposition, the Green function can be written as \cite{Bonetti2022}
\begin{equation} \label{eq: G decomp}
 \widetilde{G}(\bk,ik_0) =
 \frac{1}{2} \sum_{\ell=\pm} \frac{u_\bk^\ell}{ik_0 - E_\bk^\ell} \, ,
\end{equation}
with the matrix coefficients
\begin{equation}
 u_\bk^\ell = \sigma^0 + \ell \frac{h_\bk}{e_\bk} \sigma^3 +
 \ell \frac{\Delta_m}{e_\bk} \sigma^1 \, ,
\end{equation}
where $h_\bk = \frac{1}{2}(\xi_\bk - \xi_{\bk+\bQ})$ and
$e_\bk = \sqrt{h_\bk^2 + \Delta_m^2}$.


\subsubsection{Chargon susceptibility} \label{sec: chargon susceptibility}

The dynamical pseudospin susceptibility of the chargons is defined as the connected expectation value
\begin{equation} \label{eq: chi}
 \chi_{jj'}^{ab}(\tau) = \bra S_j^{\psi a}(\tau) S_{j'}^{\psi b}(0) \ket_c =
 \bra S_j^{\psi a}(\tau) S_{j'}^{\psi b}(0) \ket -
 \bra S_j^{\psi a}(\tau) \ket \bra S_{j'}^{\psi b}(0) \ket  \, ,
\end{equation}
with the pseudospin fields from Eq.~\eqref{eq: Sj}.
We evaluate the spin susceptibility in random phase approximation (RPA), which is a conserving approximation consistent with the mean-field treatment of the free energy and the single-particle Green function.
The spin susceptibility is thereby coupled to the charge susceptibility \cite{Kampf1996}.
Hence, we extend the definition Eq.~\eqref{eq: chi} to a combined spin-charge susceptibility by including the local charge density as additional component
$S_j^{\psi 0}(\tau) =
\frac{1}{2} \sum_{s,s'} \psi_{js}^*(\tau) \sigma^0_{ss'} \psi_{js'}(\tau)$,
with the $2\times2$ identity matrix $\sigma^0$.
We keep the factor $\frac{1}{2}$ in the charge component to be consistent with the prefactor of the spin components.

We denote the Fourier transform of $\chi_{jj'}^{ab}(\tau)$ by $\chi^{ab}(\bq,\bq',iq_0)$, and its analytic continuation to real frequencies $iq_0 \to \omega + i0^+$ by $\chi^{ab}(\bq,\bq',\omega)$. Due to the broken translation invariance in the spiral state, the susceptibility has non-zero components not only for $\bq' = \bq$, but also for $\bq' = \bq \pm \bQ$ and $\bq' = \bq \pm 2\bQ$.
The susceptibility in the rotated spin basis
$\tilde\chi_{jj'}^{ab}(\tau) =
 \bra \tilde{S}_j^{\psi a}(\tau) \tilde{S}_{j'}^{\psi b}(0) \ket_c$, with
$\tilde{S}_j^{\psi a}(\tau) =
 \frac{1}{2} \sum_{s,s'} \tilde\psi_{js}^*(\tau) \sg_{ss'}^a \tilde\psi_{js'}(\tau)$,
is instead translation invariant, so that its Fourier transform $\tilde\chi^{ab}(\bq,\omega)$ is diagonal in momentum.
For a spiral state aligned in the $xy$ plane, the diagonal components of $\chi^{ab}(\bq,\bq',\omega)$, for $a=b$ and $\bq=\bq'$, are related to $\tilde\chi^{ab}(\bq,\omega)$ by \cite{Kampf1996,Bonetti2022}
\begin{eqnarray} \label{eq: trafo dia}
 \chi^{00}(\bq,\omega) &=& \tilde\chi^{00}(\bq,\omega) ,  \\
 \chi^{11}(\bq,\omega) &=& \chi^{22}(\bq,\omega) \label{eq: chi22 unr} \\
 &=& \frac{1}{4} \big[ \tilde\chi^{11}(\bq\!+\!\bQ,\omega) +
 \tilde\chi^{11}(\bq\!-\!\bQ,\omega) \nonumber \\
 &&+ \tilde\chi^{22}(\bq\!+\!\bQ,\omega) + \tilde\chi^{22}(\bq\!-\!\bQ,\omega) \nonumber \\
 &&+ 2i \tilde\chi^{12}(\bq\!+\!\bQ,\omega) + 2i\tilde\chi^{21}(\bq\!-\!\bQ,\omega) \big] ,
 \hskip 5mm \nonumber \\
 \chi^{33}(\bq,\omega) &=& \tilde\chi^{33}(\bq,\omega) \, .
\end{eqnarray}
In a N\'eel state aligned in the $x$ direction, the relations for $\chi^{00}(\bq,\omega)$ and $\chi^{33}(\bq,\omega)$ remain the same, while
\begin{eqnarray} \label{eq: trafo dia Neel}
 \chi^{11}(\bq,\omega) &=& \tilde\chi^{11}(\bq\!+\!\bQ,\omega) \, , \\
 \chi^{22}(\bq,\omega) &=& \tilde\chi^{22}(\bq\!+\!\bQ,\omega) \, .
\end{eqnarray}
Note that $\bq + \bQ = \bq - \bQ$ for $\bQ = (\pi,\pi)$.
We also note that the above relations for the N\'eel state do not correspond to the relations for the spiral case with $\bQ = (\pi,\pi)$. This is because contributions to $\chi^{aa}(\bq,\bq \pm 2\bQ)$ for spiral order turn into additional contributions to $\chi^{aa}(\bq,\bq)$ in the N\'eel limit $\bQ \to (\pi,\pi)$ \cite{Bonetti2022}.

In renormalized RPA, the rotated susceptibility matrix composed of the matrix elements
$\tilde\chi^{ab}(\bq,\omega)$ is given by \cite{Bonetti2022, Bonetti2022a}
\begin{equation} \label{eq: rpa}
 \tilde\chi(\bq,\omega) = \tilde\chi_0(\bq,\omega)
 \left[ I_4 - \Gamma_0 \tilde\chi_0(\bq,\omega) \right]^{-1} \, ,
\end{equation}
where $I_4$ is the $4\times4$ identity matrix, and
$\Gamma_0 = 2 {\rm diag}(-\bar U,\bar U,\bar U,\bar U)$ is the interaction vertex with the renormalized interaction $\bar U$.
The bare susceptibility matrix $\tilde\chi_0(\bq,\omega)$ is composed of the elements
\begin{eqnarray}
 \tilde\chi_0^{ab}(\bq,\omega) &=&
 - \frac{1}{4} \int_\bk T \sum_{k_0} \tr \big[ \sigma^a \widetilde{G}(\bk,ik_0) \nonumber \\
 && \times \sigma^b \widetilde{G}(\bk+\bq,ik_0 + iq_0) \big]
 \Big|_{iq_0 \rightarrow \omega + i0^+} \, .
\end{eqnarray}
Using the decomposition \eqref{eq: G decomp}, the Matsubara sum over $k_0$ can be carried out, yielding \cite{Bonetti2022}
\begin{equation} \label{eq: chi0t}
 \tilde \chi_0^{ab}(\bq,\omega) =
 -\frac{1}{8} \sum_{\ell,\ell'=\pm} \int_\bk \mathcal{A}^{ab}_{\ell\ell'}(\bk,\bq)
 \frac{f(E_\bk^\ell) - f(E_{\bk+\bq}^{\ell'})}
 {\omega + i0^+ + E_\bk^\ell - E_{\bk+\bq}^{\ell'}} \, ,
\end{equation}
with the coefficients
\begin{equation}
 \mathcal{A}^{ab}_{\ell\ell'}(\bk,\bq) =
 \frac{1}{2} \tr \left( \sigma^a u_\bk^\ell \sigma^b u_{\bk+\bq}^{\ell'} \right) \, .
\end{equation}
%


\subsection{Spin stiffnesses}

In this section we provide the formulae for the pseudospin stiffnesses in the spiral state of the chargons, which are the key parameters entering the non-linear sigma model, Eq.~\eqref{eq: sigma model} for the spinon dynamics. Various derivations of expressions for the stiffnesses can be found in Refs.~\cite{Bonetti2022, Bonetti2022a, Bonetti2022ward, Goremykin2024}. They can be obtained either by expanding the inverse spin susceptibility around the Goldstone poles \cite{Bonetti2022}, or from a suitable low energy-momentum limit of the SU(2) gauge field response kernel \cite{Bonetti2022a, Bonetti2022ward, Goremykin2024}.

The stiffness matrix $\cJ_{\mu\nu}$ is diagonal in the spin indices. For a spiral state aligned in the $xy$ plane,
\begin{equation} \label{eq: stiffness matrix}
 \cJ_{\mu\nu} = {\rm diag} \left(
 J_{\mu\nu}^\perp, J_{\mu\nu}^\perp, J_{\mu\nu}^\Box \right) ,
\end{equation}
where $J_{\mu\nu}^\Box$ and $J_{\mu\nu}^\perp$ describe the stiffness of in-plane and out-of-plane fluctuations, respectively. We denote their temporal components $J_{00}^a$ by $Z^a$, and their spatial components by $J_{\alpha\beta}^a$ with $\alpha,\beta \in \{ x,y \}$.
We ignore spatio-temporal components $J_{\alpha0}^a$ and  $J_{0\beta}^a$, which are imaginary and related to Landau damping. For the out-of-plane stiffness they actually vanish, since the out-of-plane Landau damping is of cubic order in momentum and frequency \cite{Bonetti2022}. The in-plane Landau damping is quadratic in momentum and frequency, and depends also on the ratio between frequency and momentum, so that the coefficient of the quadratic behavior cannot be parametrized by a single number.
We now present expressions for the stiffnesses in the form obtained from an expansion of the susceptibility.

For the spatial in-plane stiffness one obtains \cite{Bonetti2022}
\begin{equation} \label{eq: Jin}
\begin{split}
 J_{\alpha\beta}^\Box &=
 - 2\Delta_m^2 \Big[ \partial^2_{q_\alpha q_\beta} \tilde{\chi}^{22}_0(\bq,0) \\
 &+ 2\sum_{a,b=0,1} \partial_{q_\alpha}\tilde{\chi}_0^{2a}(\bq,0)
 \tilde\Gamma_{\rm red}^{ab}(\bq,0)
 \partial_{q_\beta}\tilde{\chi}^{b2}_0(\bq,0) \Big] \Big|_{\bq=\mathbf{0}} \, ,
\end{split}
\end{equation}
where the matrix $\tilde\Gamma_{\rm red}(q)$ represents the RPA effective interaction in the subspace spanned by the charge channel and the spin channel in $x$-direction,
\begin{eqnarray}
 \tilde\Gamma_{\rm red}(q) &=&
 \left[ I_2 -
 \left( \begin{array}{cc} -2{\bar U} & 0 \\ 0 & 2{\bar U} \end{array} \right)
 \left( \begin{array}{cc} \tilde{\chi}_0^{00}(q) & \tilde{\chi}_0^{01}(q) \\
 \tilde{\chi}_0^{10}(q) & \tilde{\chi}_0^{11}(q) \end{array} \right)
 \right]^{-1} \nonumber \\
 &\times& \left( \begin{array}{cc} -2{\bar U} & 0 \\ 0 & 2{\bar U} \end{array} \right) \, ,
\end{eqnarray}
with $q = (\bq,\omega)$. The matrix elements $\tilde\Gamma_{\rm red}^{ab}(q)$ with $a,b \in \{0,1\}$ are all finite for $\omega = 0$ and $\bq \to \mathbf{0}$.
More explicit expressions for the momentum derivatives of the bare susceptibilities appearing in Eq.~\eqref{eq: Jin} are listed in App.~\ref{app: spin stiffnesses}.

For the spatial out-of-plane stiffness one obtains \cite{Bonetti2022, Bonetti2022ward, factor2}
\begin{equation} \label{eq: Jout}
 J^{\perp}_{\alpha\beta} = - \left. \Delta_m^2 \partial^2_{q_\alpha q_\beta}
 \tilde{\chi}^{33}_0(\bq,0) \right|_{\bq=\bQ} \, .
\end{equation}
It is fully determined by the bare susceptibility. A more explicit expression is provided in App.~\ref{app: spin stiffnesses}.
For a spiral state with a wave vector of the form $\bQ = (\pi-2\pi\eta,\pi)$, the spatial spin stiffnesses are diagonal in the spatial indices, that is, $J_{xy}^a = J_{yx}^a = 0$, while $J_{xx}^a \neq J_{yy}^a$ for $\eta > 0$.

The temporal in-plane stiffness can be written as \cite{Bonetti2022}
\begin{eqnarray} \label{eq: Zin}
 Z^\Box &=& 2\Delta_m^2 \Big[
 \partial_\omega^2 \tilde\chi_0^{22}(\mathbf{0},\omega) \big|_{\omega=0} \nonumber \\
 &+& \frac{4 \bar U}{1 - 2 \bar U \tilde\chi_0^{33}(\mathbf{0},\omega \to 0)}
 \left|\partial_\omega \tilde\chi_0^{23}(\mathbf{0},\omega)\big|_{\omega=0}\right|^2
 \Big] \, .
\end{eqnarray}
The temporal out-of-plane stiffness can be written as \cite{Bonetti2022, Bonetti2022ward, factor2}
\begin{equation} \label{eq: Zout}
 Z^\perp = \Delta_m^2 \Big[
 \left. \partial_\omega^2 \tilde\chi_0^{33}(\bQ,\omega) \right|_{\omega=0}
 + \sum_{a,b \in \{0,1,2\}}
 \left. \partial_\omega \tilde\chi_0^{3a}(\bQ,\omega) \right|_{\omega=0}
 \tilde\Gam^{ab}(\bQ,0)
 \left. \partial_\omega \tilde\chi_0^{b3}(\bQ,\omega) \right|_{\omega=0}
 \Big] \, ,
\end{equation}
where $\tilde\Gam^{ab}$ is the RPA effective interaction in the rotated spin frame.
Expressions for the frequency derivatives of the bare susceptibilities appearing in Eqs.~ \eqref{eq: Zin} and \eqref{eq: Zout} are provided in App.~\ref{app: spin stiffnesses}.

For a N\'eel state with spins aligned in the $x$ direction, the stiffness matrix has the form
\begin{equation} \label{eq: stiffness matrix Neel}
 \cJ_{\mu\nu} = {\rm diag} \left( 0, J_{\mu\nu}, J_{\mu\nu} \right) ,
\end{equation}
with $J_{\alpha\beta} = J \delta_{\alpha\beta}$. The stiffnesses $J$ and $Z = J_{00}$ can be obtained from the above general formulae for $Z^a$ and $J_{\alpha\beta}^a$ in the spiral state by setting $\bQ = (\pi,\pi)$, where the expressions for the out-of-plane stiffnesses need to be multiplied by two \cite{Bonetti2022, Bonetti2022ward, factor2}.
$J$ is obtained most easily from Eq.~\eqref{eq: Jout}, with an extra factor of two, and $Z$ from Eq.~\eqref{eq: Zin}.


\section{Electron spin susceptibility} \label{sec: electron spin}

In this section we derive expressions for the spin susceptibility of the electrons
\begin{equation} \label{eq: spin suscept}
 S_{jj'}^{ab}(\tau) = \bra S_j^a(\tau) S_{j'}^b(0) \ket_c =
 \bra S_j^a(\tau) S_{j'}^b(0) \ket \, ,
\end{equation}
where $S_j^a(\tau) = \frac{1}{2} c_j^*(\tau) \sigma^a c_j(\tau)$ is now the spin field formed from the \emph{electron} fields (instead of chargons).
We consider only the thermally or quantum disordered case, where the disconnected term
$\bra S_j^a(\tau) \ket \bra S_{j'}^b(0) \ket$ vanishes.


\subsection{Fractionalization}

Inserting the electron fractionalization Eq.~\eqref{eq: electron fract} into Eq.~\eqref{eq: spin suscept}, we obtain
\begin{eqnarray}
 S_{jj'}^{ab}(\tau) &=&
 \frac{1}{4} \bra \psi_j^*(\tau) R_j^\dag(\tau) \sigma^a R_j(\tau) \psi_j(\tau)
 \psi_{j'}^*(0) R_{j'}^\dag(0) \sigma^b R_{j'}(0) \psi_{j'}(0) \ket \nonumber \\
 &=& \frac{1}{4} \bra \psi_j^*(\tau) \cR_j^{ac}(\tau) \sigma^c \psi_j(\tau)
 \psi_{j'}^*(0) \cR_{j'}^{bd}(0) \sigma^d \psi_{j'}(0) \ket \, ,
\end{eqnarray}
where in the second step we have passed to the adjoint representation Eq.~\eqref{eq: adjoint rep}. Neglecting correlations between chargons and spinons, we factorize the above expectation value in chargon and spinon expectation values as
\begin{equation} \label{eq: suscept fac}
 S_{jj'}^{ab}(\tau) = \bra \cR_j^{ac}(\tau) \cR_{j'}^{bd}(0) \ket
 \bar\chi_{jj'}^{cd}(\tau) \, ,
\end{equation}
where $\bar\chi_{jj'}(\tau)$ is the pseudospin susceptibility of the chargons as defined in Eq.~\eqref{eq: chi}, but with the disconnected contributions included.
An analogous factorization approximation has been used in calculations of the single-electron propagator \cite{Scheurer2018, Wu2018, Bonetti2022a}.


\subsection{Spinon correlator}

We now evaluate the spinon correlator in Eq.~\eqref{eq: suscept fac},
\begin{equation} \label{eq: def D}
 \cD_{jj'}^{acbd}(\tau) = \bra \cR_j^{ac}(\tau) \cR_{j'}^{bd}(0) \ket \, ,
\end{equation}
where, in the continuum limit, the expectation value is determined by the spinon action Eq.~\eqref{eq: sigma model}.
The SO(3) matrix $\cR$ can be expressed as a triad of orthonormal unit vectors,
\begin{equation} \label{eq: R triad}
 \cR = \big({\bf\hat n}^1,{\bf\hat n}^2,{\bf\hat n}^3 \big),
\end{equation}
which are subject to the local constraints
\begin{equation} \label{eq: constraints}
 {\bf\hat n}_j^c(\tau) \cdot {\bf\hat n}_j^d(\tau) = \delta_{cd} \, .
\end{equation}
Note that the index labeling the three unit vectors corresponds to the column index of the matrix $\cR$.
For a spin-diagonal stiffness matrix $\cJ_{\mu\nu}$, the spinon action in Eq.~\eqref{eq: sigma model} can then be written as
\begin{equation} \label{eq: S[R]}
 \cS[\cR] = \frac{1}{2} \int \! dx \sum_c P_{\mu\nu}^{cc} \,
 [\partial_\mu {\bf\hat n}^c(x)] \cdot \partial_\nu {\bf\hat n}^c(x) \, ,
\end{equation}
with Einstein's summation convention for the repeated space-time indices $\mu$ and $\nu$.

Inserting the stiffness matrix \eqref{eq: stiffness matrix} of a spiral state into the relation
$\cP_{\mu\nu} = \frac{1}{2} \tr(\cJ_{\mu\nu}) \, I_3 - \cJ_{\mu\nu}$, we obtain the relations
\begin{equation} \label{eq: P_spiral}
 P_{\mu\nu}^{11} = P_{\mu\nu}^{22} = {\textstyle \frac{1}{2}} J_{\mu\nu}^\Box \, , \;
 P_{\mu\nu}^{33} = J_{\mu\nu}^\perp - {\textstyle \frac{1}{2}} J_{\mu\nu}^\Box \, .
\end{equation}
For the N\'eel state, Eq.~\eqref{eq: stiffness matrix Neel} yields
\begin{equation} \label{eq: P_Neel}
 P_{\mu\nu}^{11} = J_{\mu\nu} \, , \;
 P_{\mu\nu}^{22} = P_{\mu\nu}^{33} = 0 \, .
\end{equation}

In frequency-momentum representation, the spinon correlator defined in Eq.~\eqref{eq: def D} is given by
\begin{equation} \label{eq: Dacbd(q)}
 \cD^{acbd}(q) = \bra \hat n_a^c(q) \hat n_b^d(-q) \ket \, .
\end{equation}
Using the CP$^1$ representation of the non-linear sigma model, we show in Appendix~\ref{app: CP1} that the propagator has the diagonal form
\begin{equation}
 \cD^{acbd}(q) = \delta_{ab} \delta_{cd} \, \cD^c(q) \, .
\end{equation}
Neglecting the current-current interaction in the CP$^1$ action, which can be justified in a large $N$ limit, all spinon correlators are equal:
$\cD^1 = \cD^2 = \cD^3 \equiv \cD$. This is also shown in Appendix~\ref{app: CP1}.
The electron spin susceptibility from Eq.~\eqref{eq: suscept fac} can then be written in the form
\begin{equation} \label{eq: spin structure1}
 S_{jj'}^{ab}(\tau) =
 \cD_{jj'}(\tau) \left[ \bar\chi_{jj'}^{11}(\tau) + \bar\chi_{jj'}^{22}(\tau) +
 \bar\chi_{jj'}^{33}(\tau) \right] \delta_{ab} \, ,
\end{equation}
where
\begin{equation}
 \cD_{jj'}(\tau) = \int_q \cD(q) \, e^{i\bq\cdot(\br_j-\br_{j'})} e^{-iq_0 \tau} \, .
\end{equation}
Inserting the spin expectation value of the chargons, Eq.~\eqref{eq: spiral}, into the disconnected contributions to $\bar\chi^{aa}(\tau)$, we obtain
\begin{subequations} \label{eq: chi bar chi}
\begin{align}
 & \bar\chi_{jj'}^{11}(\tau) =
 \chi_{jj'}^{11}(\tau) + m^2 \cos(\bQ\cdot\br_j) \cos(\bQ\cdot\br_{j'}) \\
 & \bar\chi_{jj'}^{22}(\tau) =
 \chi_{jj'}^{22}(\tau) + m^2 \sin(\bQ\cdot\br_j) \sin(\bQ\cdot\br_{j'}) \\
 & \bar\chi_{jj'}^{33}(\tau) = \chi_{jj'}^{33}(\tau) \, .
\end{align}
\end{subequations}
Using
$\cos(\bQ\cdot\br_j) \cos(\bQ\cdot\br_{j'}) + \sin(\bQ\cdot\br_j) \sin(\bQ\cdot\br_{j'}) =
 \cos[\bQ\cdot(\br_j - \br_{j'})]$, Eq.~\eqref{eq: spin structure1} yields
\begin{equation} \label{eq: spin structure2}
 S_{jj'}^{ab}(\tau) =
 \cD_{jj'}(\tau) \left[ m^2 \cos[\bQ\cdot(\br_j - \br_{j'})] +
 \tr \, \chi_{jj'}(\tau) \right] \delta_{ab} \, ,
\end{equation}
with $\tr \, \chi_{jj'}(\tau) =
\chi_{jj'}^{11}(\tau) + \chi_{jj'}^{22}(\tau) + \chi_{jj'}^{33}(\tau)$.
This function is manifestly spin SU(2) and translation invariant. The translation invariance follows from the translation invariance of the chargon susceptibility $\tilde\chi_{jj'}$ in the rotated spin basis and the invariance of the trace under a change of basis.
The SU(2) symmetry of the spinons thus restores the broken symmetries of the chargon susceptibility.

We now evaluate the spinon correlator $\cD(q)$ in a saddle point approximation, which becomes exact in a suitable large $N$ limit.
We begin with the N\'eel case, where the action $\cS[\cR]$ depends only on ${\bf\hat n}^1$,
\begin{equation} \label{eq: S[R] Neel}
 \cS[\cR] = \frac{1}{2} \int dx \, J_{\mu\nu} \,
 [\partial_\mu {\bf\hat n}^1(x)] \cdot \partial_\nu {\bf\hat n}^1(x) \, .
\end{equation}
The dynamics of ${\bf\hat n}^2$ and ${\bf\hat n}^3$ is exclusively determined by the orthonormality constraints.
The propagator for ${\bf\hat n}^1$ is fully determined by the action Eq.~\eqref{eq: S[R] Neel} and the normalization ${\bf\hat n}^1(x) \cdot {\bf\hat n}^1(x) = 1$. Implementing the normalization constraint by a Lagrange multiplier field $\lam(x)$, we obtain the action
\begin{equation} \label{eq: S[R,lam] Neel}
 \cS[\cR,\lam] = \frac{1}{2} \int dx \left\{
 J_{\mu\nu} \, [\partial_\mu {\bf\hat n}^1(x)] \cdot \partial_\nu {\bf\hat n}^1(x) +
 i\lam(x) \left[ {\bf\hat n}^1(x) \cdot {\bf\hat n}^1(x) - 1 \right] \right\} \, .
\end{equation}
In a saddle point approximation we replace the fluctuating field $\lam(x)$ by a space-time independent number $\lam$. The propagator for ${\bf\hat n}^1$ can then be read off directly from the action as
\begin{equation}
 \bra \hat n_a^1(q) \hat n_b^1(-q) \ket =  \delta_{ab} \cD^1(q) =
 \frac{\delta_{ab}}{Z (m_s^2 + q_0^2) + J \bq^2} \, ,
\end{equation}
with $m_s^2 = i\lam/Z$. Since $\cD^1 = \cD^2 = \cD^3 \equiv \cD$, we have found
\begin{equation} \label{eq: cD Neel}
 \cD(q) = \frac{1}{Z (m_s^2 + q_0^2) + J \bq^2} \, .
\end{equation}
The dynamically generated spinon mass $m_s$ is determined by the normalization condition
$\bra {\bf\hat n}^1(x) \cdot {\bf\hat n}^1(x) \ket = 1$, corresponding to
\begin{equation}
 \int_q \cD(q) = \int_q \frac{1}{Z (m_s^2 + q_0^2) + J \bq^2} = \frac{1}{3} \, .
\end{equation}
The above integral is ultraviolet divergent. It thus requires an ultraviolet cutoff which we will specify below.

We now turn to the case of spiral order. Neglecting the current-current interaction in the CP$^1$ representation \eqref{eq: CP1 action} of the non-linear sigma model, as before,
the action becomes independent of $J^\Box$. Hence, setting $J^\Box = 0$ doesn't affect the results obtained from the CP$^1$ representation, and the saddle point solution for the spiral case has the same form as for the N\'eel case.
The action \eqref{eq: S[R]} obtained from the adjoint representation of the rotation matrices can also be brought to the N\'eel form by setting $J^\Box = 0$, such that
\begin{equation} \label{eq: S[R] spiral}
 \cS[\cR] = \frac{1}{2} \int dx \, J_{\mu\nu}^\perp \,
 [\partial_\mu {\bf\hat n}^3(x)] \cdot \partial_\nu {\bf\hat n}^3(x) \, .
\end{equation}
This action has the same form as the action for the N\'eel case, Eq.~\eqref{eq: S[R] Neel}, with ${\bf\hat n}^1$ replaced by ${\bf\hat n}^3$. It depends only on the out-of-plane stiffness $J^\perp$. Applying the same saddle point approximation as for the N\'eel case above, we obtain the spinon correlator
$\cD = \cD^1 = \cD^2 = \cD^3$ as
\begin{equation} \label{eq: cD spiral}
 \cD(q) =
 \frac{1}{Z^\perp(m_s^2 + q_0^2) + J_{\alpha\beta}^\perp q_\alpha q_\beta} \, ,
\end{equation}
where the spinon mass $m_s$ is again determined by the condition $\int_q \cD(q) = \frac{1}{3}$.

The spinon correlators in the N\'eel and in the spiral state have thus the same form. The same is true for the spinon propagator $D(q) = \bra z_\sigma^*(q) z_\sigma(q) \ket$ on the saddle point level in the CP$^1$ representation \cite{Bonetti2022a}, see Appendix~\ref{app: CP1 saddle}.

Fourier transforming Eq.~\eqref{eq: spin structure2} yields the electron spin susceptibility as a function of momentum and (Matsubara) energy, $S^{ab}(q) = \delta_{ab} S(q)$, with
\begin{eqnarray} \label{eq: spin structure3}
 S(\bq,iq_0) &=&
 \frac{1}{2} m^2 \left[ \cD(\bq+\bQ,iq_0) + \cD(\bq-\bQ,iq_0) \right] \nonumber \\
 &+& T \sum_{q'_0} \int_{\bq'}
 \cD(\bq',iq'_0) \, \tr\chi(\bq-\bq',iq_0-iq'_0) \, .
\end{eqnarray}
%


\section{Results} \label{sec: results}

\begin{figure}[tb]
\centering
\includegraphics[width=0.6\linewidth]{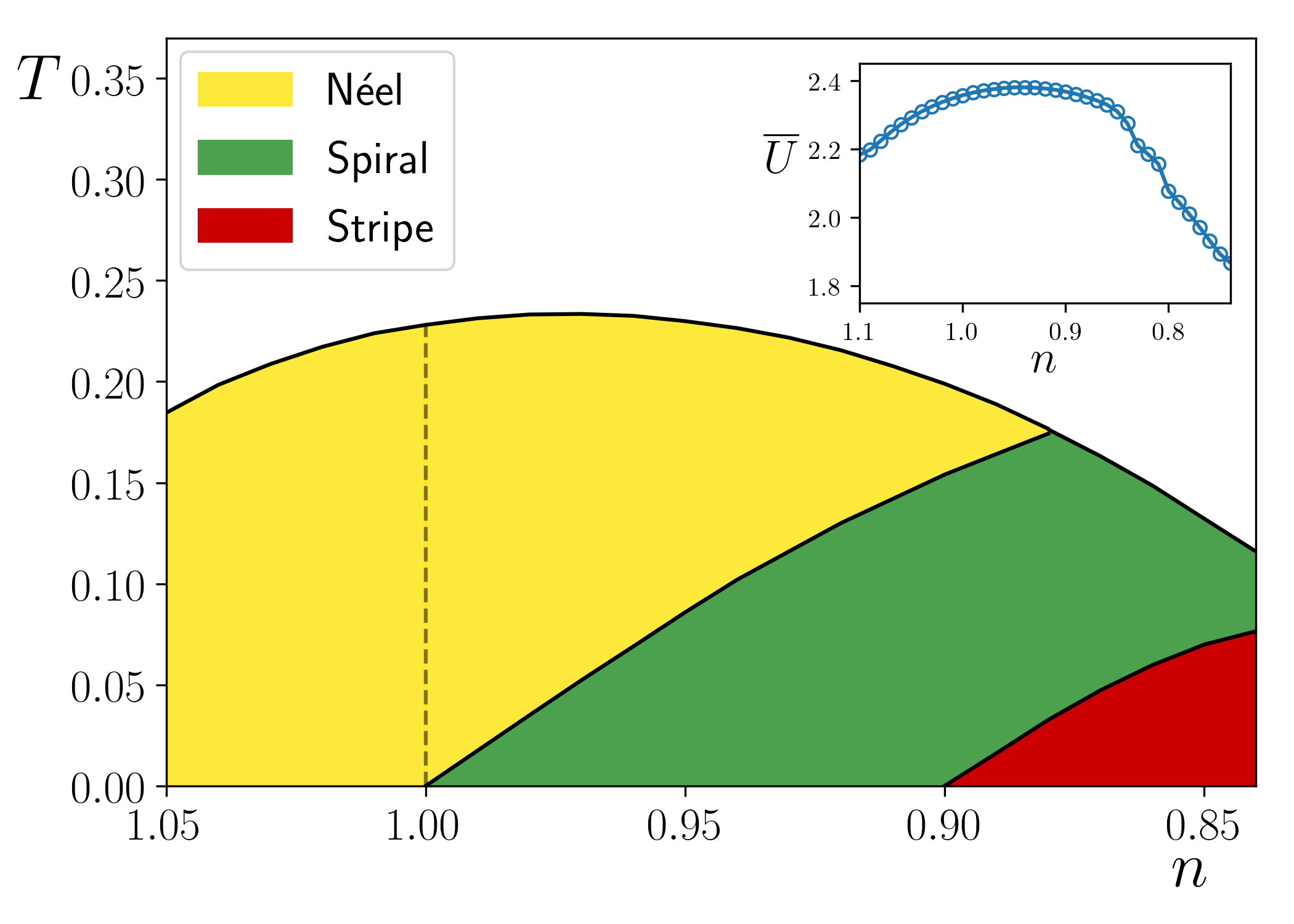}
 \caption{Chargon phase diagram with N\'eel, spiral, and spin-charge ordered (stripe) states for densities $n$ around half-filling. The inset shows the effective interaction $\bar U$ as a function of the density.}
\label{fig: phasedia}
\end{figure}
We now present explicit results for the spin susceptibility. All results are obtained for the Hubbard model on a square lattice with $t' = -0.2t$ and $U = 4t$. We choose the nearest-neighbor hopping amplitude $t$ as our unit of energy.
In Fig.~\ref{fig: phasedia} we show the chargon phase diagram for these parameters \cite{Robin}, as obtained from the renormalized mean-field theory described in Sec.~\ref{sec: chargons}. Qualitatively, the phase diagram has the same structure as magnetic phase diagrams obtained in Hartree-Fock theory for a sizable $t'$ at $U = 3t$ \cite{Scholle2023, Scholle2024}. The region labeled by ``stripe'' is characterized by spin-charge stripe order in a broad sense, where the spin order is not necessarily collinear. Non-collinear states with a superposition of many wave vectors appear especially in a small region close to the spiral regime \cite{Scholle2024}.
In the following we will evaluate the spin susceptibility exclusively for parameters where the chargons exhibit N\'eel or circular spiral order.


\subsection{Chargon pseudospin susceptibility}

To understand the results for the electron spin susceptibility, it is useful to discuss properties of the pseudospin susceptibility of the chargons first. We need to consider three qualitatively distinct cases: N\'eel order at half-filling, N\'eel order away from half-filling, and spiral order. The latter occurs only in the hole doped regime below half-filling.
We focus on the diagonal components $\chi^{aa}(\bq,\omega)$ of the general chargon susceptibility $\chi^{ab}(\bq,\bq',\omega)$, since off-diagonal elements (in spin or momentum) do not contribute to the electron susceptibility when the SU(2) symmetry and translation symmetry are fully restored by the spinon fluctuations.

\begin{figure}[tb]
\centering
\includegraphics[width=0.49\linewidth]{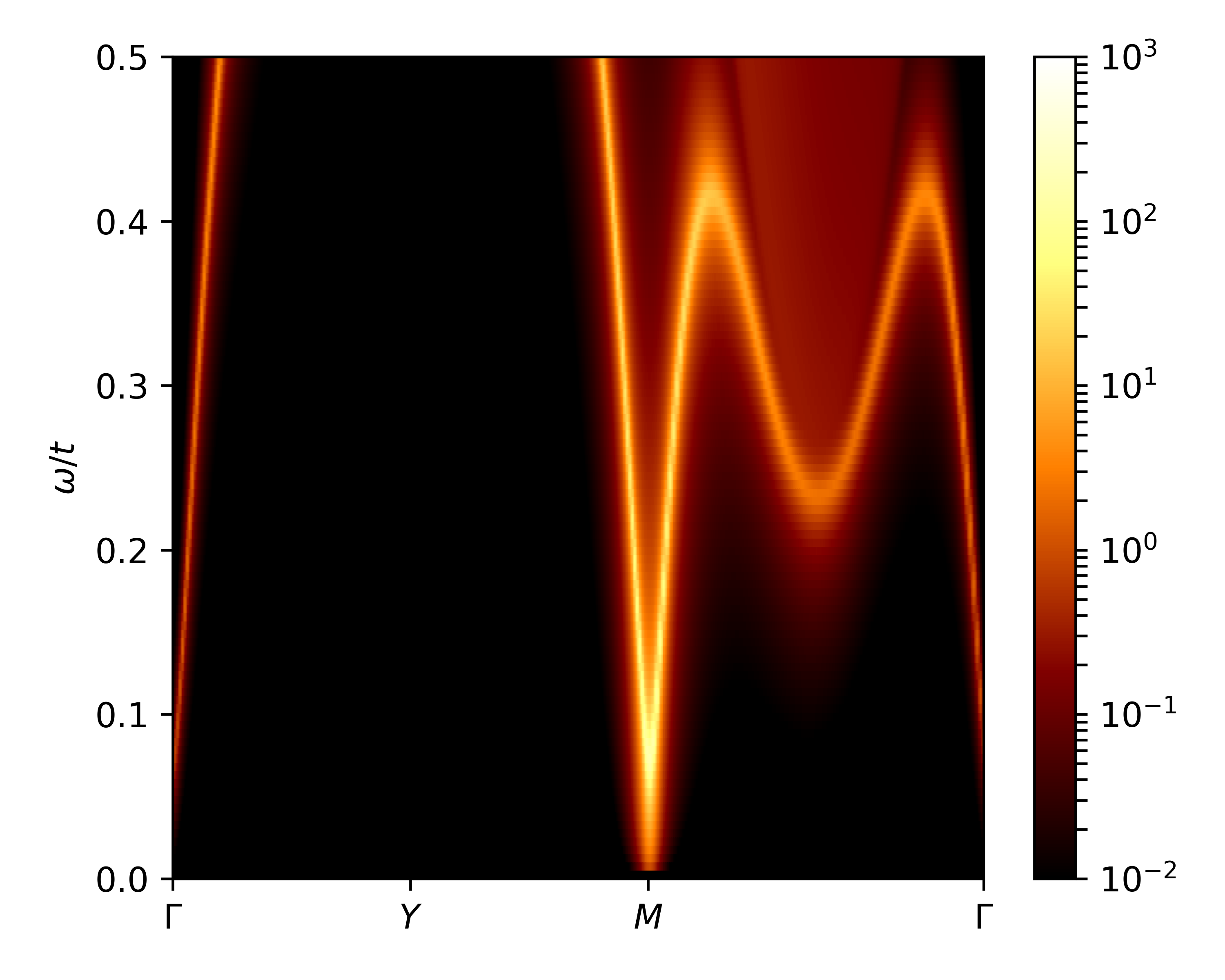}
 \caption{${\rm Im}\chi^{22}(\bq,\omega) = {\rm Im} \chi^{33}(\bq,\omega)$ in the N\'eel state at half-filling along the path $\Gamma \to Y \to M \to \Gamma$ in momentum space.
 The temperature is $T=0.05t$.}
\label{fig: Im chi 1}
\end{figure}
\begin{figure}[tb]
\centering
\includegraphics[width=0.5\linewidth]{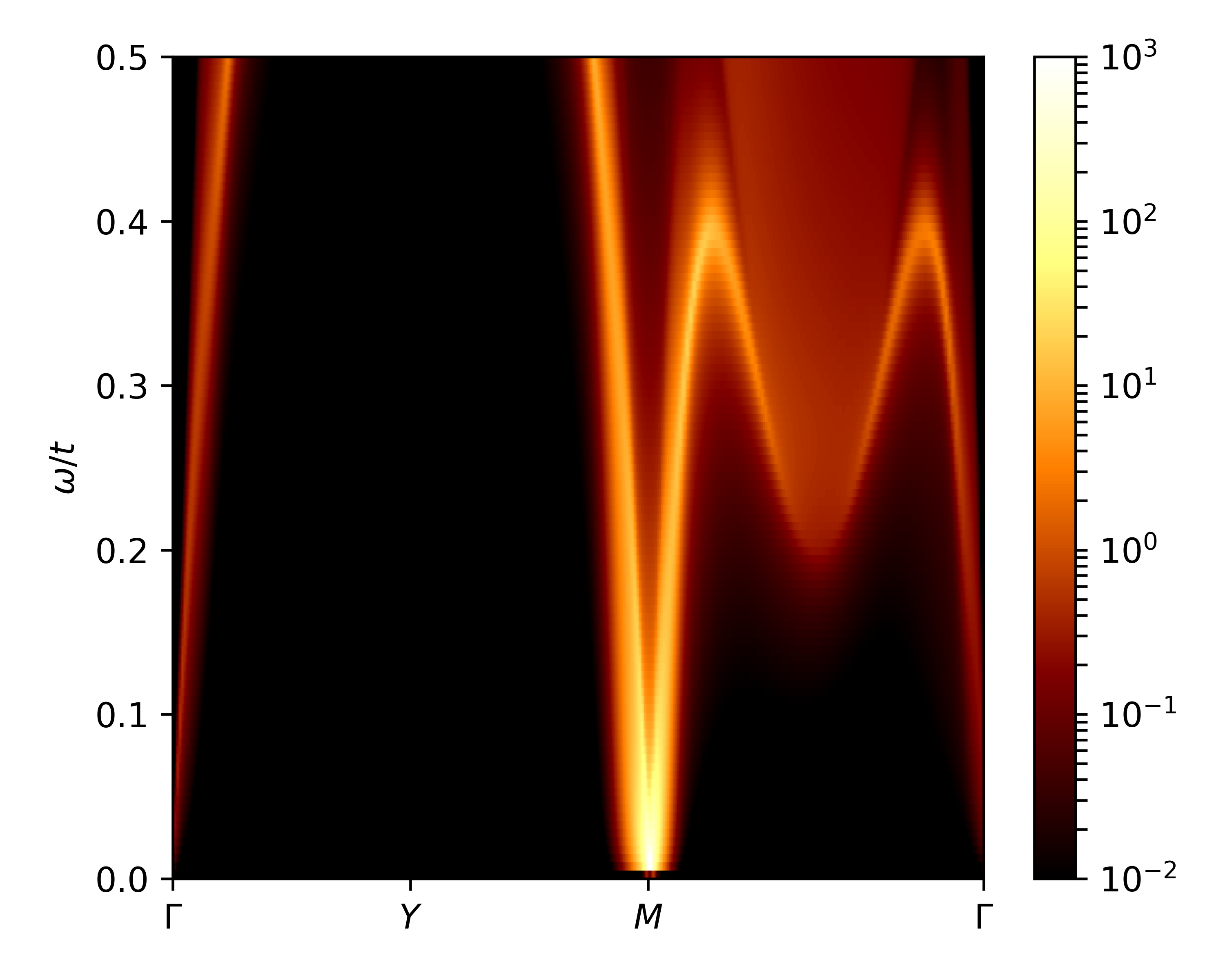}
 \caption{${\rm Im}\chi^{22}(\bq,\omega) = {\rm Im} \chi^{33}(\bq,\omega)$ in the N\'eel state at $n = 0.98$ and $T = 0.05t$ along the path $\Gamma \to Y \to M \to \Gamma$ in momentum space.}
\label{fig: Im chi 2}
\end{figure}
\begin{figure}[tb]
\centering
\includegraphics[width=0.49\linewidth]{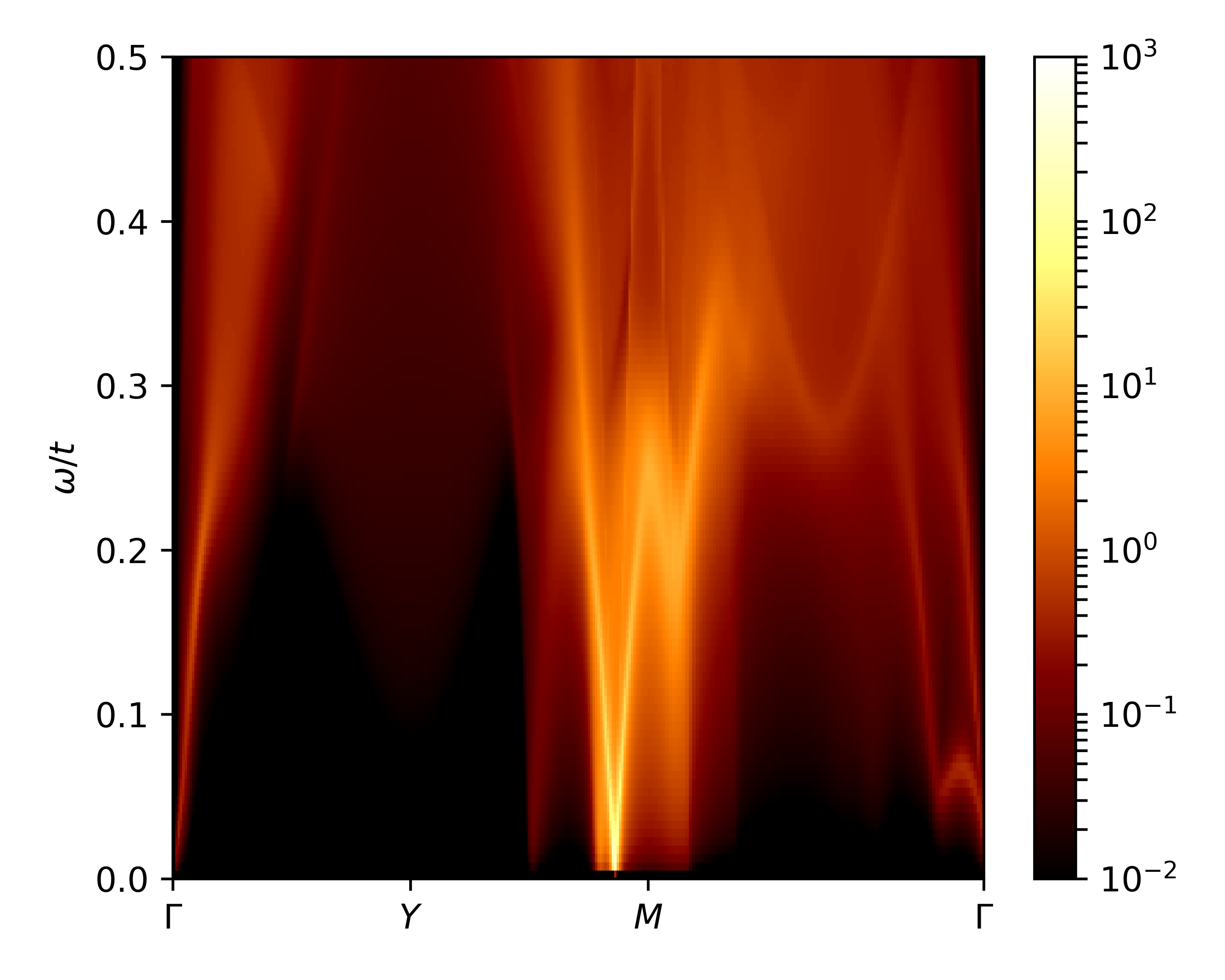}
\includegraphics[width=0.49\linewidth]{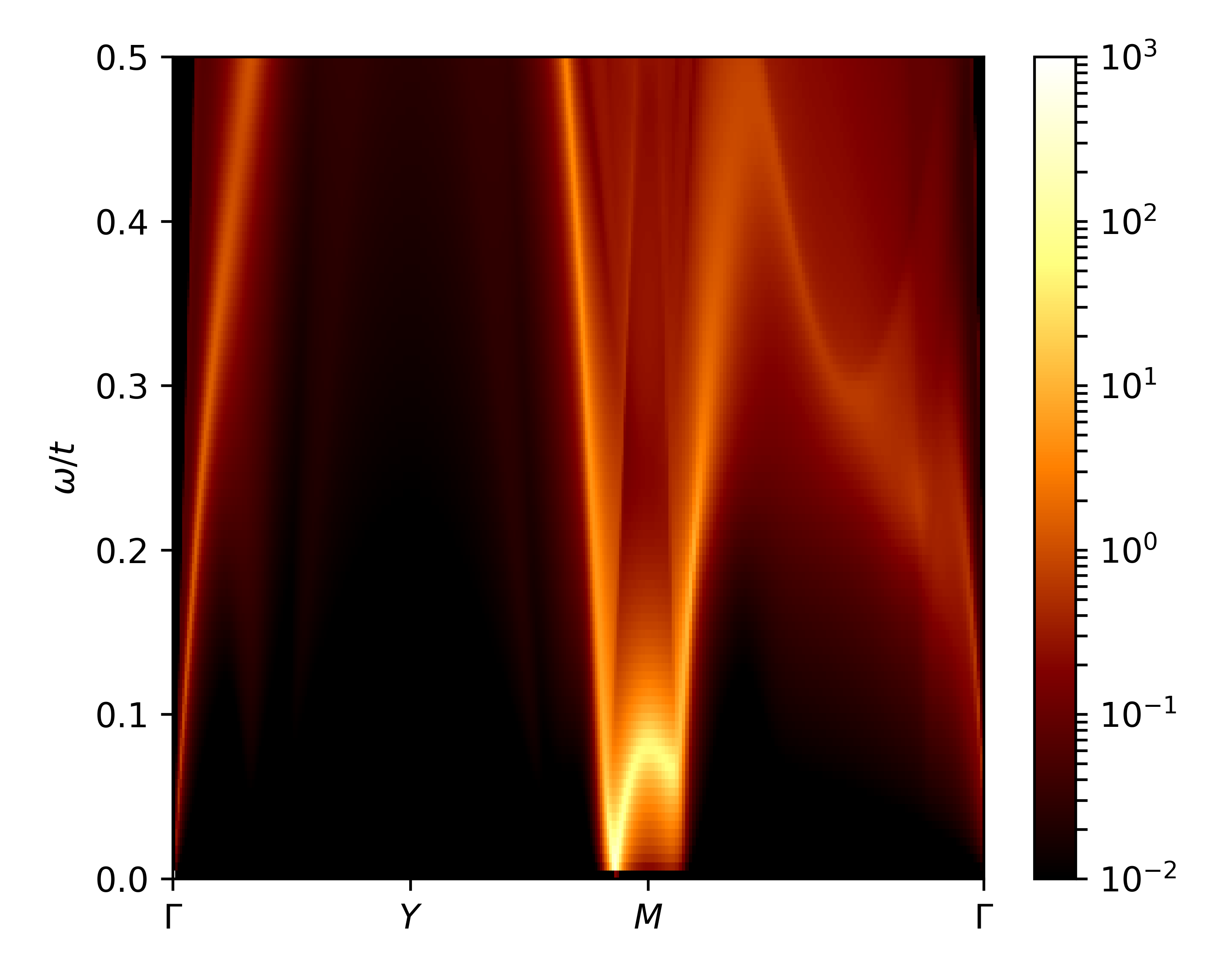}
 \caption{${\rm Im}\chi^{11}(\bq,\omega) = {\rm Im}\chi^{22}(\bq,\omega)$ (left) and
 ${\rm Im} \chi^{33}(\bq,\omega)$ (right) in the spiral state at $n = 0.90$ and $T = 0.02t$ along the path $\Gamma \to Y \to M \to \Gamma$ in momentum space.}
\label{fig: Im chi 3}
\end{figure}

A global overview of the spin excitations can be obtained by looking at the imaginary part of $\chi^{aa}(\bq,\omega)$ along a momentum path in the Brillouin zone. In Figs.~\ref{fig: Im chi 1} - \ref{fig: Im chi 3} we plot ${\rm Im}\chi^{aa}(\bq,\omega)$ for various densities along the path $\Gamma \to Y \to M \to \Gamma$ with
$\Gamma = (0,0)$, $Y = (0,\pi)$, and $M = (\pi,\pi)$.
We do not show the longitudinal component ${\rm Im}\chi^{11}(\bq,\omega)$ in the N\'eel state, since this quantity is completely structureless and its intensity remains very low.
In all plots one can see branches of spin excitations with a well defined energy-momentum relation $\Omega_\bq$. These are collective modes associated with poles of $\chi^{aa}(\bq,\omega)$, that is, with zeros of the determinant
$\det[I_4 - \Gamma_0 \tilde\chi_0(\bq,\omega)]$.
In particular, the Goldstone modes are clearly visible in $\chi^{22} = \chi^{33}$ near the $M$ point in the N\'eel state, and in $\chi^{11} = \chi^{22}$ and $\chi^{33}$ near the ordering wave vector $\bQ = (\pi - 2\pi\eta,\pi)$ in the spiral state. Their dispersion is linear and their spectral weight remains finite in the low energy limit. In the spiral state, the dispersions of the in-plane and out-of-plane modes, visible in $\chi^{11} = \chi^{22}$ and $\chi^{33}$, respectively, have distinct slopes as expected. The properties of the Goldstone modes in the spiral state, that is, their dispersion, their spectral weight, and their Landau damping have been discussed in detail in Ref.~\cite{Bonetti2022}.
There are also bands of low energy spin excitations with a small momentum, that is, near the $\Gamma$-point, but their spectral weight vanishes in the low energy limit.
The collective modes are particularly sharp at half-filling, due to the absence of Landau damping in this case. There is very little broadening due to the finite temperature in the shown plots, that is, the same quantities at lower temperatures look very much the same.

Besides the collective modes associated with the poles of $\chi^{aa}(\bq,\omega)$, there is also a rather featureless continuum of spin excitations. For $T \to 0$ at half-filling they are exclusively due to interband contributions to $\tilde\chi_0^{ab}$, and are thus confined to high energies. Away from half-filling intraband contributions lead to a continuum of excitations also at low energies, but their spectral weight is so low that this is not visible in the plots -- in spite of the logarithmic intensity scale.

\begin{figure}[tb]
\centering
\includegraphics[width=0.49\linewidth]{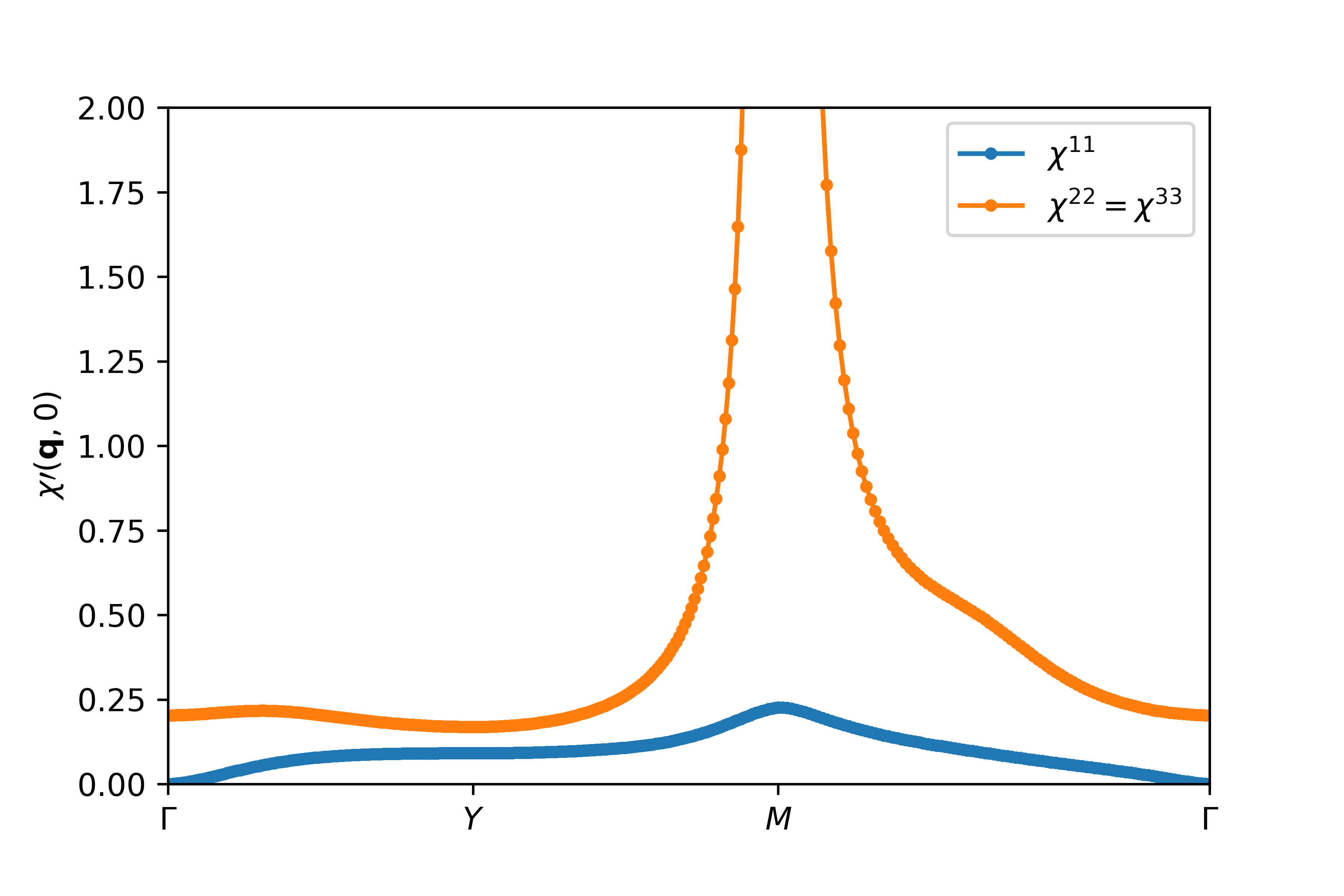}
\includegraphics[width=0.49\linewidth]{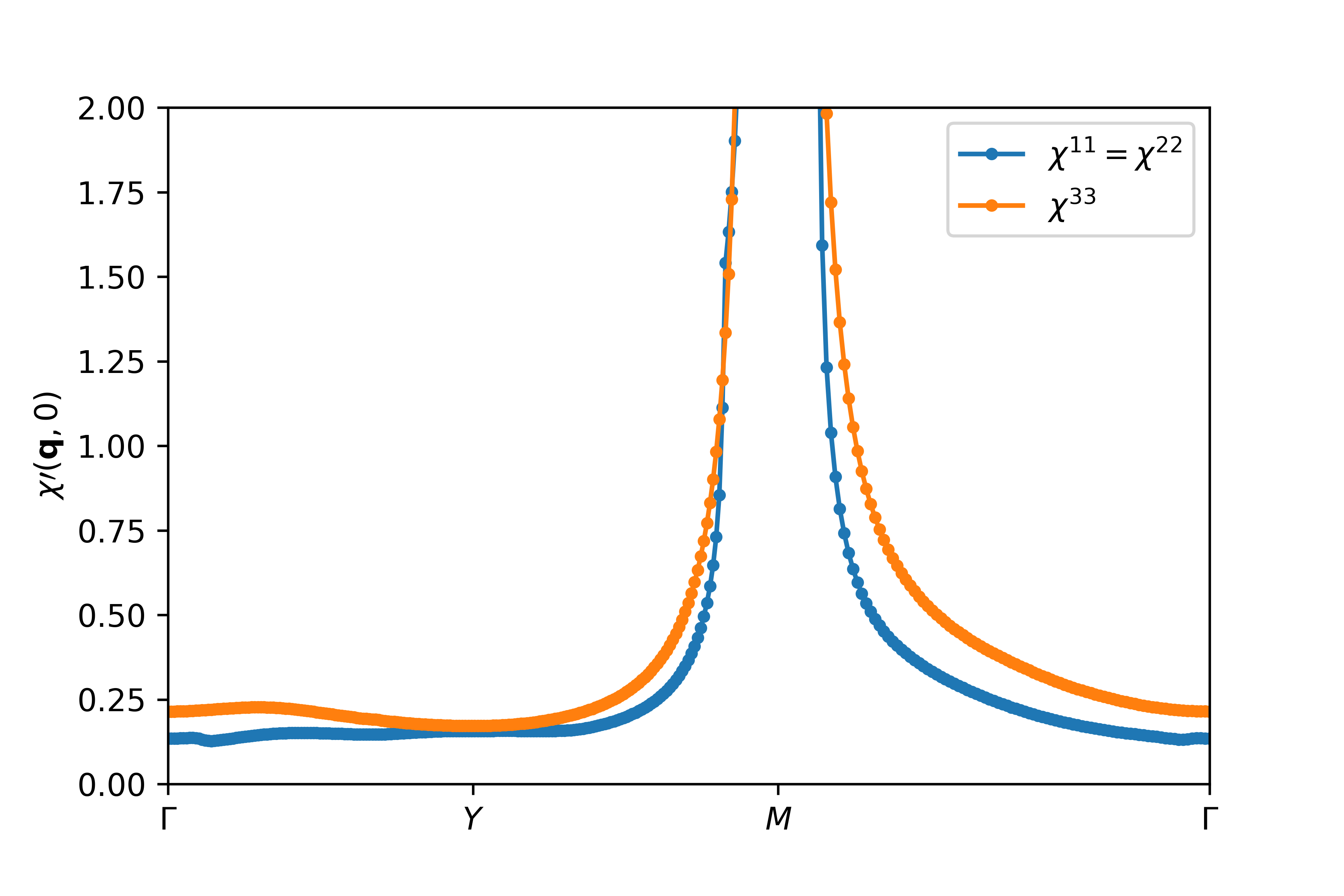}
 \caption{$\chi^{11}(\bq,0)$ and $\chi^{22}(\bq,0) = \chi^{33}(\bq,0)$ in the N\'eel state at half-filling (left), and $\chi^{11}(\bq,0) = \chi^{22}(\bq,0)$ and $\chi^{33}(\bq,0)$ in the spiral state at $n = 0.90$ (right) along the path $\Gamma \to Y \to M \to \Gamma$.
 The temperature is T = 0.0015t in both plots.}
\label{fig: chi static}
\end{figure}
The static susceptibilities $\chi^{aa}(\bq,0)$ are real.
For $\omega=0$ the sector $a=3$ decouples from the charge and the other spin sectors \cite{Bonetti2022}, so that the static out-of-plane susceptibility $\chi^{33}(\bq,0)$ is given by the simple formula
\begin{equation}
 \chi^{33}(\bq,0) =
 \frac{\tilde\chi_0^{33}(\bq,0)}{1 - 2 \bar U \tilde\chi_0^{33}(\bq,0)} \, .
\end{equation}
In the N\'eel state, the sector $a=2$ also decouples from the rest, and the same equation holds for both transverse susceptibilities $\chi^{22} = \chi^{33}$.
In Fig.~\ref{fig: chi static} we show $\chi^{aa}(\bq,0)$ in the N\'eel state at half-filling and in a spiral state at $n = 0.90$ along the path $\Gamma \to Y \to M \to \Gamma$ in momentum space for a very low temperature. Except for the longitudinal component $\chi^{11}(\bq,0)$ in the N\'eel state, one can see that the static susceptibilities are very large in a broad region around the ordering wave vector, compared to their values far away from it. This is due to the divergence for $\bq \to \pm \bQ$ associated with the Goldstone mode.

\begin{figure}[tb]
\centering
\includegraphics[width=0.5\linewidth]{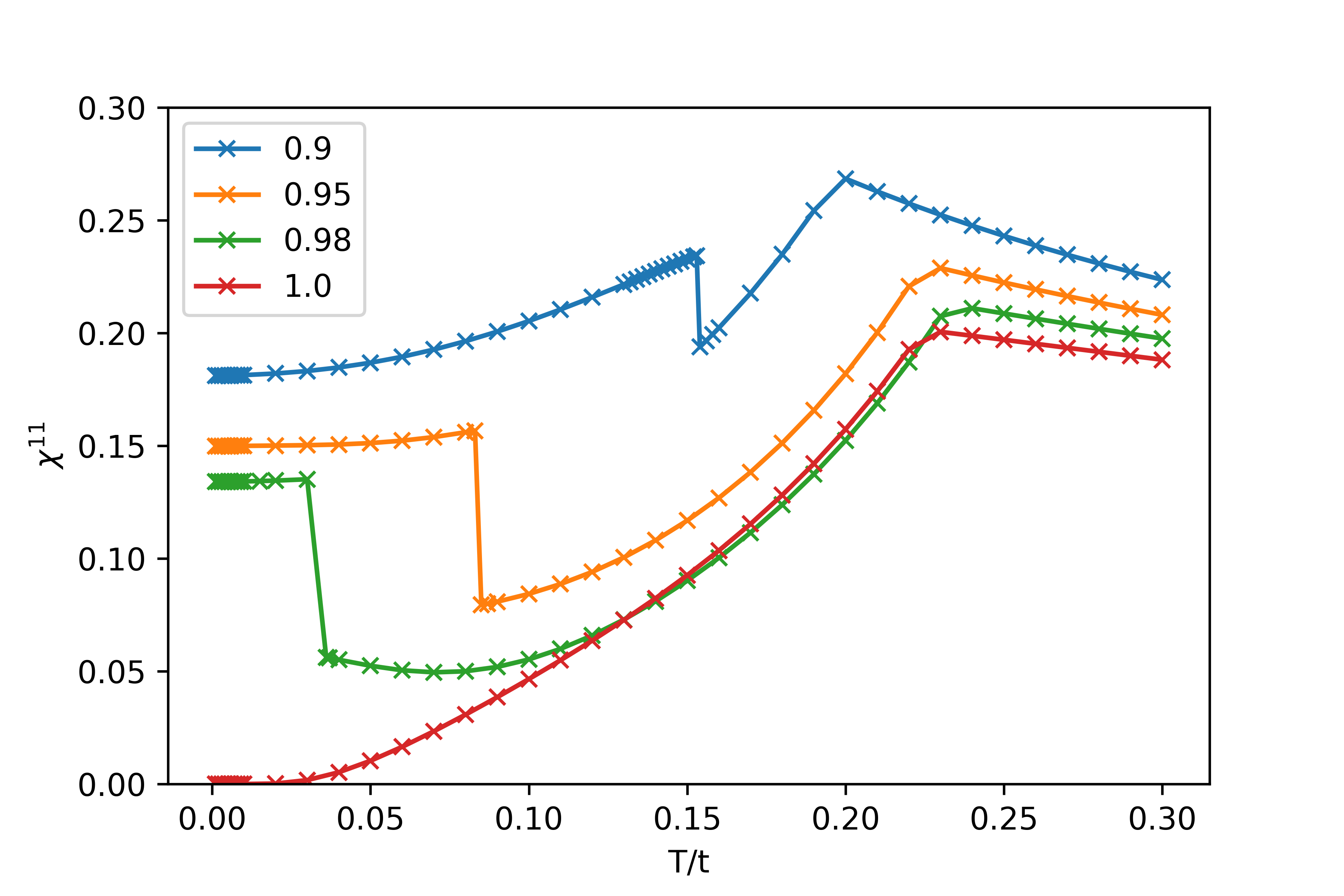}
\includegraphics[width=0.5\linewidth]{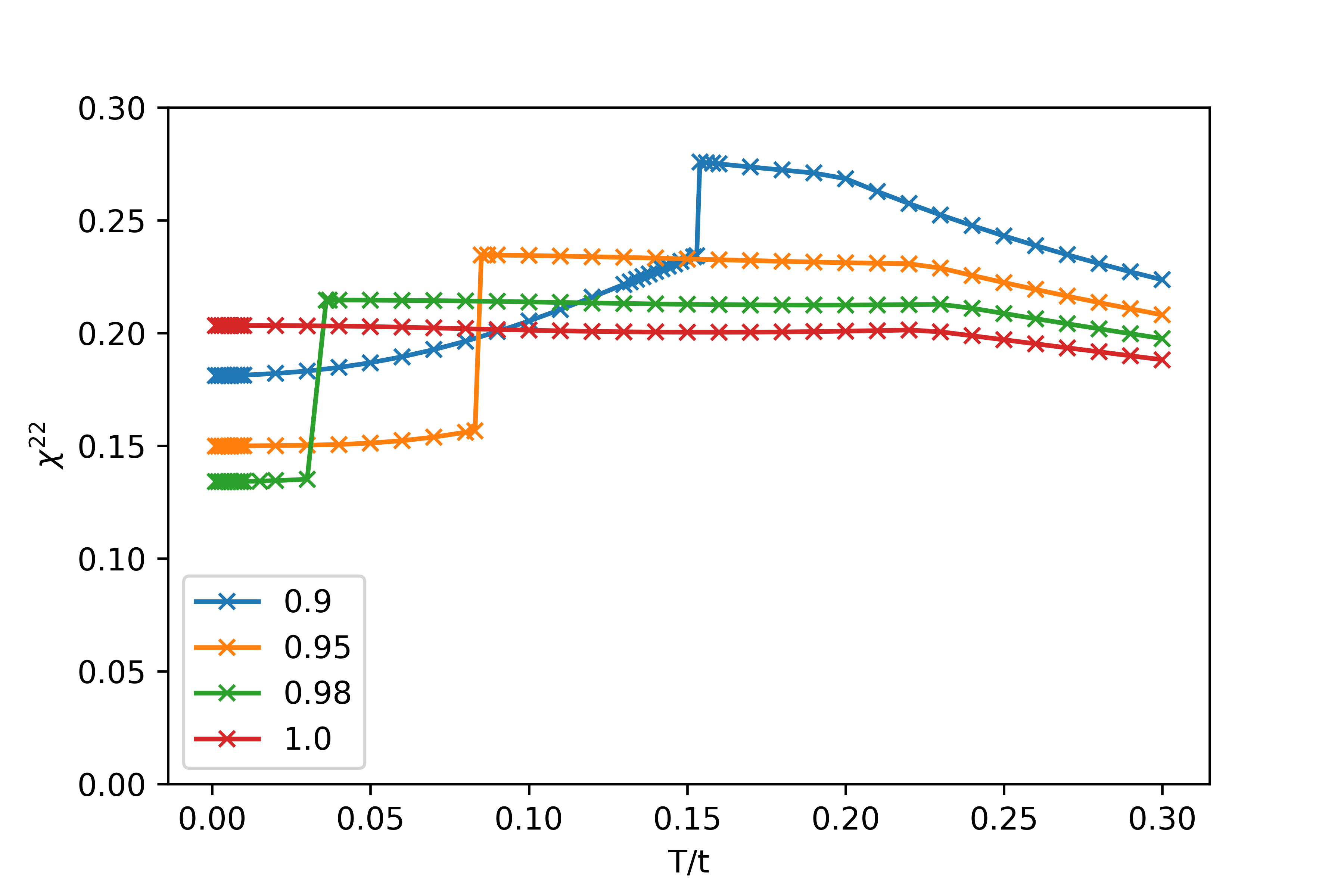}
\includegraphics[width=0.5\linewidth]{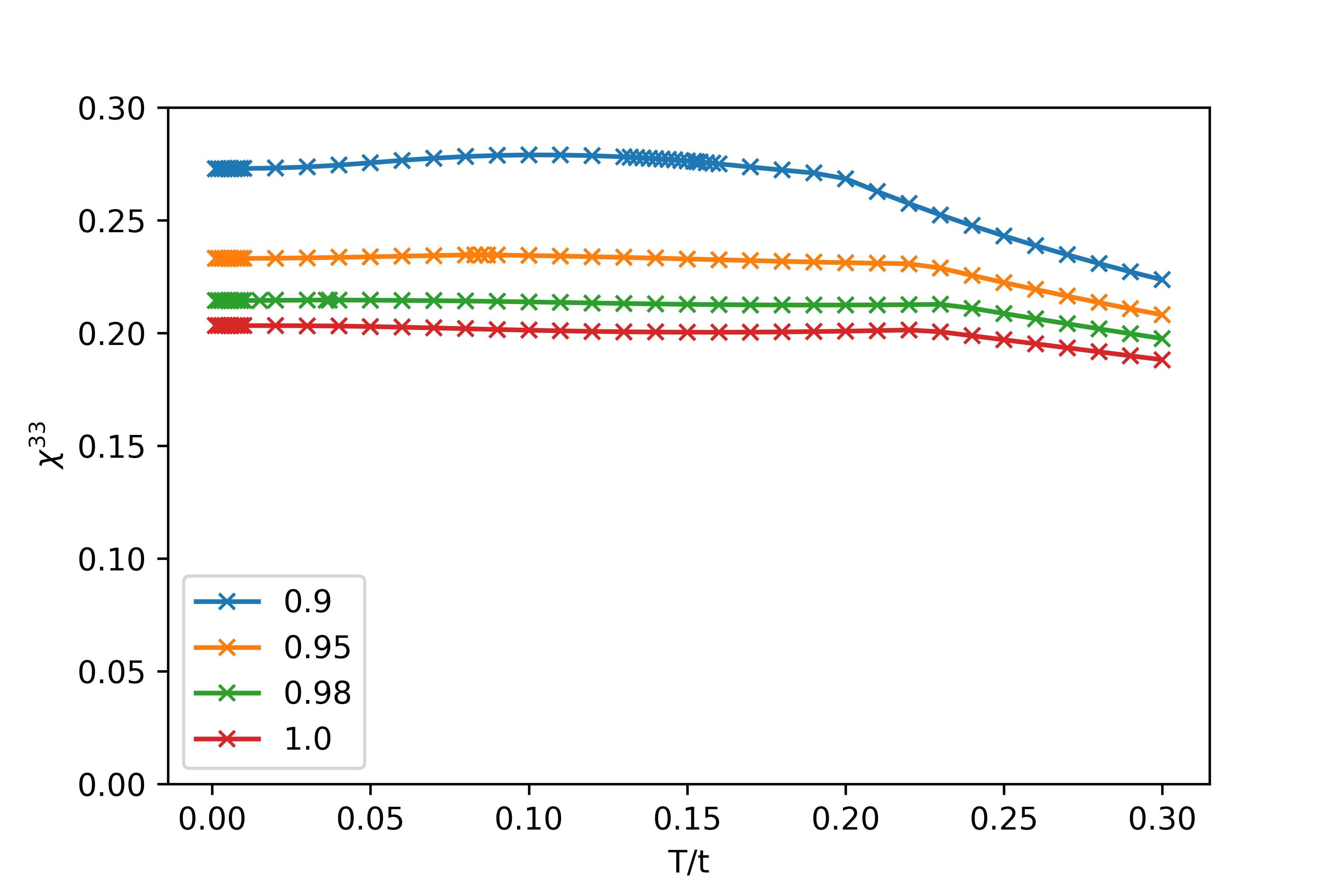}
 \caption{Uniform static chargon susceptibilities $\chi_u^{11}$, $\chi_u^{22}$, and $\chi_u^{33}$ as functions of temperature for various densities.}
\label{fig: chi_u}
\end{figure}
\begin{figure}[tb]
\centering
\includegraphics[width=0.5\linewidth]{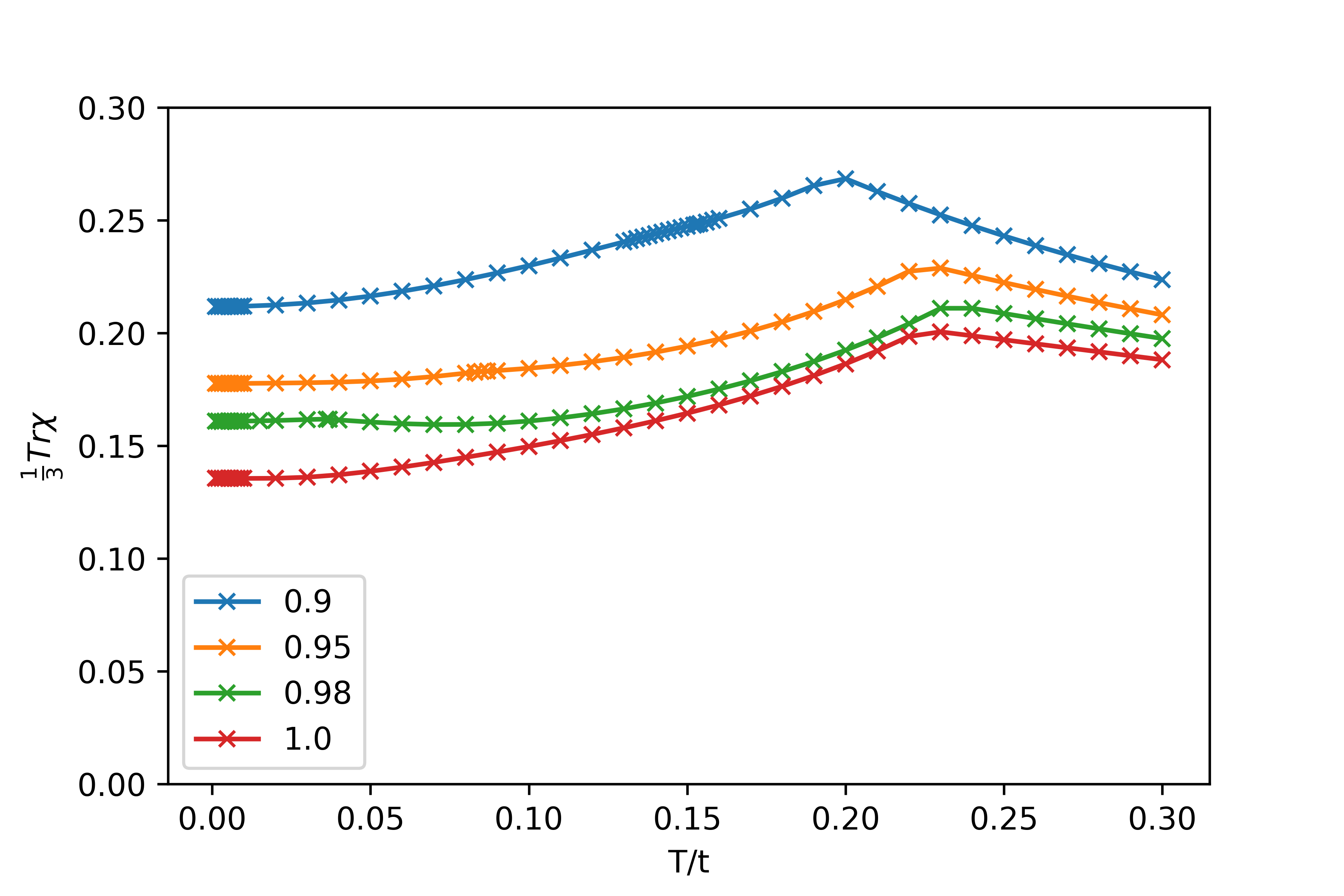}
 \caption{Average susceptibility $\frac{1}{3 }\tr\chi_u =
 \frac{1}{3} \left(\chi_u^{11} + \chi_u^{22} + \chi_u^{33}\right)$ as function of temperature for various densities.}
\label{fig: trchi_u}
\end{figure}
The uniform static susceptibilities are defined by
$\chi_u^{aa} = \chi^{aa}(\bq\to\mathbf{0},0)$.
They describe the linear spin response to a uniform external magnetic field in $a$-direction.
In Figs.~\ref{fig: chi_u} we plot the temperature dependence of the uniform static susceptibilities of the chargons at various densities.
In the N\'eel state at half-filling, the longitudinal component $\chi_u^{11}$ vanishes exponentially for $T \to 0$, while the transverse components $\chi_u^{22} = \chi_u^{33}$ are almost temperature independent for $T < T^*$ and finite.
For $n < 1$ all components $\chi_u^{aa}$ remain finite for $T \to 0$.
There are jumps in $\chi_u^{11}$ and $\chi_u^{22}$ at the transition between N\'eel and spiral order. The jump goes in opposite directions for $\chi_u^{11}$ and $\chi_u^{22}$, and it cancels out in the arithmetic average $\frac{1}{3} \tr\,\chi_u = \frac{1}{3} \left(\chi_u^{11} + \chi_u^{22} + \chi_u^{33}\right)$, shown in Fig.~\ref{fig: trchi_u}, which is the quantity entering our formula for the electron spin susceptibility.
The finite values of $\chi_u^{aa}$ at $T=0$ are due to Pauli-type intraband contributions from the hole pockets, and interband contributions.
At half-filling all intraband contributions to the uniform static susceptibilities vanish exponentially for $T \to 0$, since $f'(E_\bk^\ell)$ vanishes exponentially for $E_\bk^\ell \neq 0$.


\subsection{Electron spin susceptibility}

We now evaluate the electron spin susceptibility $S(\bq,\omega)$. More specifically, we will compute and discuss the spin structure factor, the static spin susceptibility, and the NMR relaxation rate. Most expressions have the same form for the cases of N\'eel and spiral order. In those expressions, we denote the spatial and temporal out-of-plane stiffnesses in case of spiral order by the same symbols $J$ and $Z$, respectively, as the stiffnesses in the N\'eel case.

We impose an ultraviolet cutoff $\Lam$ on the support of the spinon propagators in momentum space by multiplying it with the Gaussian cutoff function $e^{-\bq^2/(2\Lam^2)}$. The choice of the cutoff function is quite arbitrary, but $\Lam$ is in any case much smaller than $|\bQ|$.
For the sake of consistency, the same cutoff has to be chosen in the evaluation of the sum rule $\int_q \cD(q) = \frac{1}{3}$, which determines the spinon mass, and for the convolution of chargons and spinons as in Eq.~\eqref{eq: spin structure3}.
The contributions from the disconnected chargon susceptibility (the first line in Eq.~\eqref{eq: spin structure3}) to ${\rm Im} S(\bq,0)$ vanish rapidly for $\bq$ outside circles of radius $\Lam$ around $\bQ$ and $-\bQ$. In particlular, they are negligible for small $\bq$.
We choose $\Lam=0.3$ in all the following results. With this choice, the ground state remains magnetically ordered at half-filling (implying $m_s=0$), and becomes quantum disordered (with $m_s > 0$) for any degree of hole doping considered in this work.


\subsubsection{Spin structure factor}

The dynamical spin structure factor, which can be measured in particular by inelastic neutron scattering, is essentially (up to a Bose factor) determined by the imaginary part of the dynamical spin susceptibility.
Continuing Eq.~\eqref{eq: spin structure3} to real frequencies we obtain
\begin{equation} \label{eq: spin structure4}
 {\rm Im} S(\bq,\omega) =
 \frac{1}{2} m^2 \, {\rm Im} \left[ \cD(\bq+\bQ,\omega) + \cD(\bq-\bQ,\omega) \right]
 + {\rm Im} S_c(\bq,\omega) \, ,
\end{equation}
with the part originating from the connected chargon susceptibility,
\begin{equation} \label{eq: spin structure5}
 {\rm Im} S_c(\bq,\omega) =
 \frac{1}{\pi} \int d\omega' \int_{\bq'}
 {\rm Im} \cD(\bq',\omega') \, {\rm Im} \, \tr\chi(\bq-\bq',\omega-\omega')
 \left[ b(\omega') - b(\omega'-\omega) \right] \, ,
\end{equation}
where $b(\omega) = \left[ e^{\omega/T} - 1 \right]^{-1}$ is the Bose function.
The imaginary part of the spinon correlator
$\cD(\bq,\omega) = \left[ Z \omega_\bq^2 - Z(\omega + i0^+)^2 \right]^{-1}$
has the form
\begin{equation}
 {\rm Im} \cD(\bq,\omega) = \frac{\pi}{2 Z \omega_\bq}
 \left[ \delta(\omega-\omega_\bq) - \delta(\omega+\omega_\bq) \right] \, ,
\end{equation}
with the spinon dispersion
\begin{equation}
\omega_\bq = \sqrt{m_s^2 + J_{\alpha\beta} q_\alpha q_\beta/Z} \, .
\end{equation}
Performing the frequency integral in Eq.~\eqref{eq: spin structure5} yields
\begin{equation} \label{eq: spin structure6}
 {\rm Im} S_c(\bq,\omega) =
 \frac{1}{2} \sum_{s=\pm} \int_{\bq'}
 \frac{s}{Z \omega_{\bq'}}
 {\rm Im} \, \tr\chi(\bq-\bq',\omega-s\omega_{\bq'})
 \left[ b(s\omega_{\bq'}) - b(s\omega_{\bq'}-\omega) \right] \, .
\end{equation}
At zero temperature the difference of Bose functions in Eq.~\eqref{eq: spin structure6} vanishes for $|\omega| < \omega_{\bq'}$, and the equation can be simplified to
\begin{equation} \label{eq: spin structure6a}
 {\rm Im} S_c(\bq,\omega) =
 \frac{1}{2} \int_{\bq'}
 \frac{1}{Z \omega_{\bq'}}
 {\rm Im} \, \tr\chi(\bq-\bq',\omega-\omega_{\bq'}) \,
 \Theta(\omega-\omega_{\bq'}) \quad \mbox{for} \quad \omega \geq 0 \, ,
\end{equation}
and ${\rm Im} S_c(\bq,-\omega) = - {\rm Im} S_c(\bq,\omega)$.
Hence, ${\rm Im} S(\bq,\omega)$ vanishes for all $|\omega| < m_s$, that is, the electron spin susceptibility inherits the {\em spin gap}\/ from the spinons.

\begin{figure}[tb]
\centering
\includegraphics[width=0.49\linewidth]{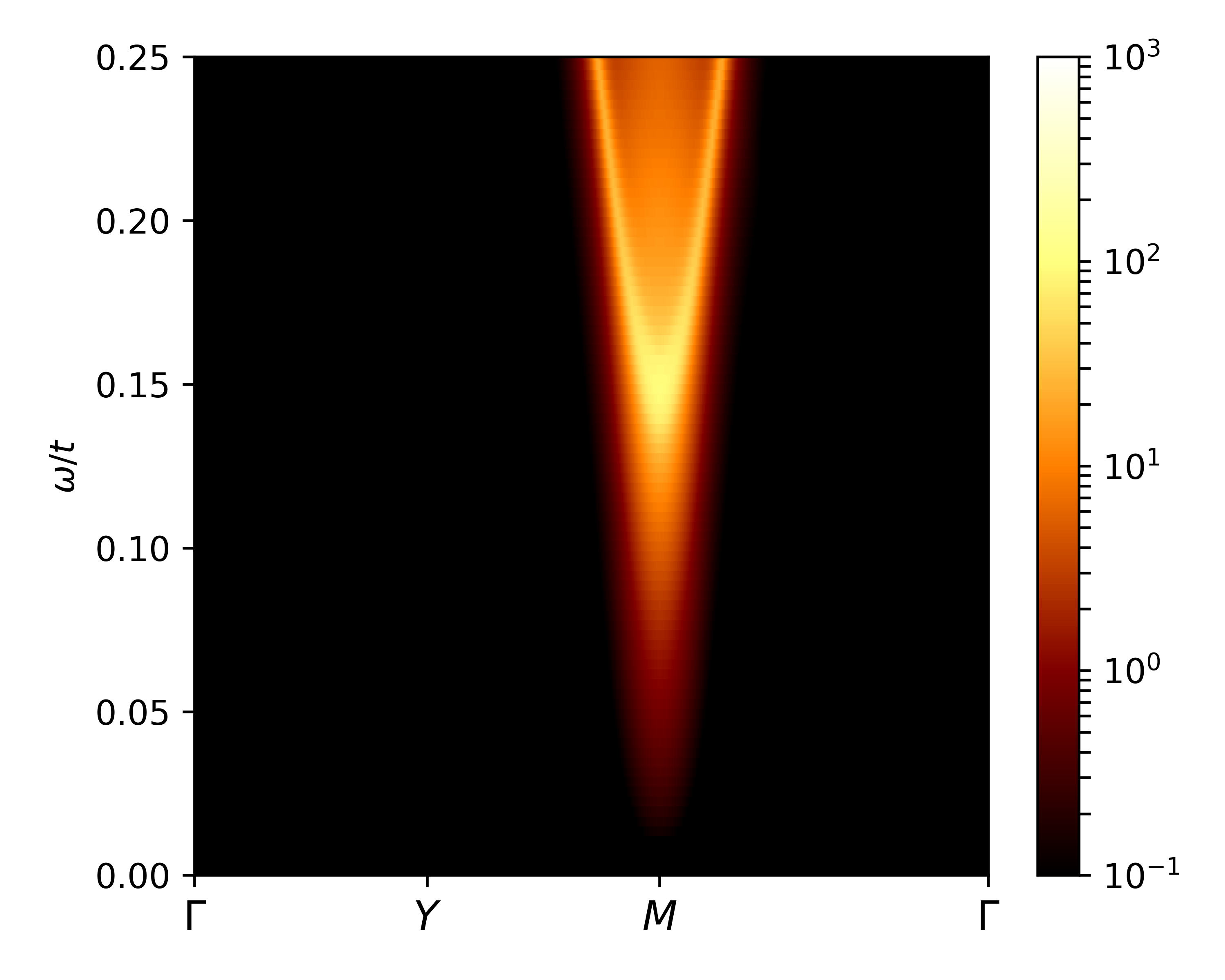}
\includegraphics[width=0.49\linewidth]{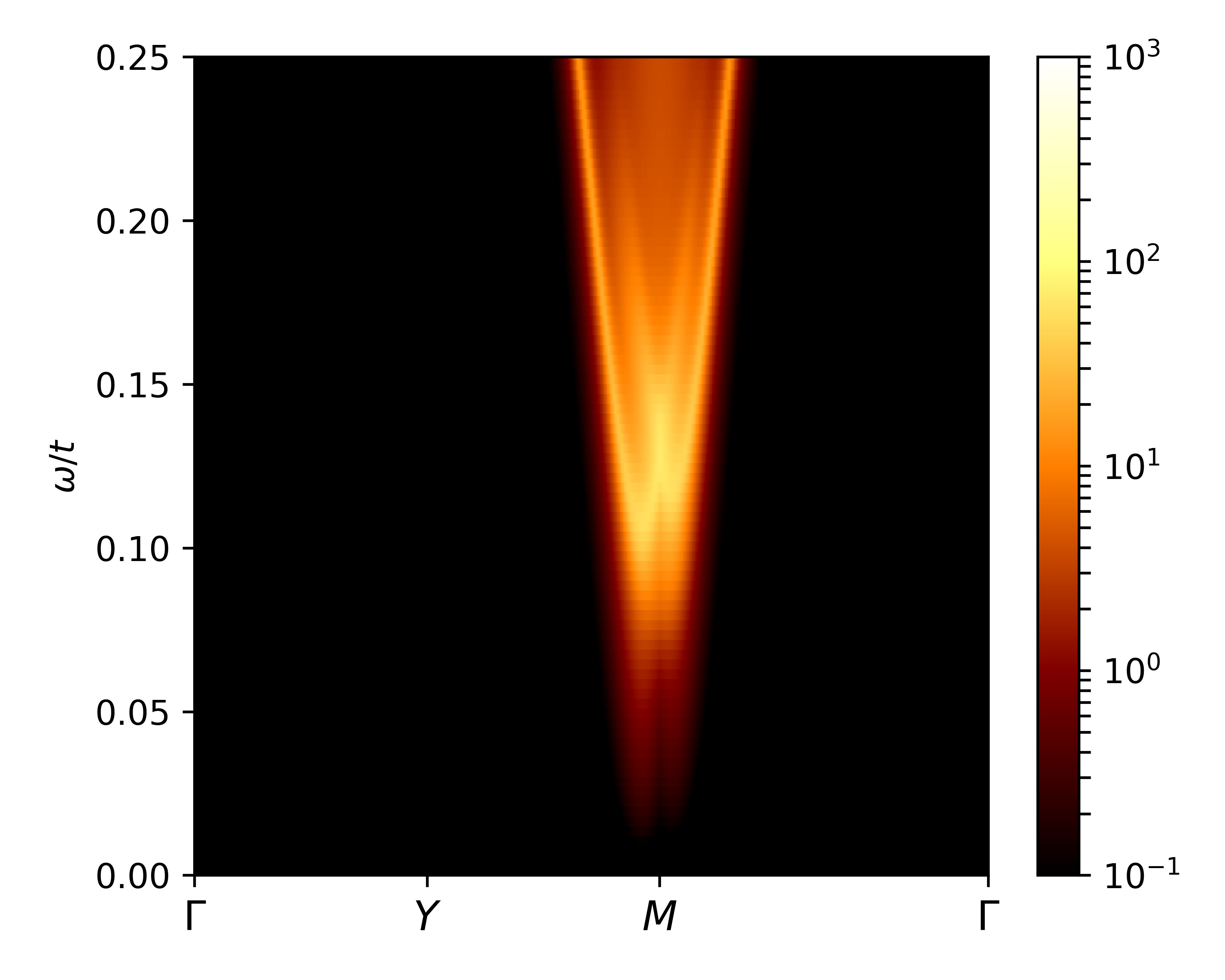}
 \caption{${\rm Im} S(\bq,\omega)$ at two distinct points in the phase diagram along the path $\Gamma \to Y \to M \to \Gamma$ in momentum space.
 Left: $n = 0.98$ and $T = 0.05$ (N\'eel state with $m_s=0.15$).
 Right: $n = 0.95$ and $T = 0.02t$ (spiral state with $\bQ=(2.9,\pi)$ and $m_s=0.11t$).
 The delta function contribution from the direct spinon term has been broadened to a Lorentzian with a damping = 0.01.}
\label{fig: Im S}
\end{figure}
In Figs.~\ref{fig: Im S} we plot ${\rm Im} S(\bq,\omega)$ at two distinct points in the phase diagram along the path $\Gamma \to Y \to M \to \Gamma$ with
$\Gamma = (0,0)$, $Y = (0,\pi)$, and $M = (\pi,\pi)$.
The underlying chargon state is N\'eel ordered in the first case (left), and spiral ordered in the second (right).
The spin gap is clearly visible.
The sharp delta function form of the direct spinon term (the first term in Eq.~\eqref{eq: spin structure4}) is an artifact of the simple saddle point solution of the spinon action. Including fluctuations around the saddle point (related to $1/N$ corrections), would endow the spinons with a decay rate, leading to a broadening of the delta function \cite{Chubukov1994}.
In the figure we have broadened the delta function by an arbitrarily chosen small damping term equal to $0.01$.
The continuous part stemming from the convolution of chargons and spinons is naturally much broader than the chargon susceptibility (see Figs.~\ref{fig: Im chi 1} - \ref{fig: Im chi 3}).
The spectral weight for $\omega < m_s$ is due to the finite temperature.
The excitation branch exhibited by the chargon susceptibility at small $\bq$ (close to the $\Gamma$ point) is also broadened by the spinon fluctuations. Its intensity is too weak to be seen in the plot.

The intensity plots in Figs.~\ref{fig: Im S} are reminiscent of the upper part of the ``hour glass'' dispersion observed via neutron scattering in the pseudogap regime of cuprate superconductors \cite{Fujita2012}. In the spiral regime (right panel) there are also two separate but fuzzy branches below the spinon gap energy, which resemble the lower part of the hour glass dispersion.
The downward branches of the hour glass dispersion in cuprates can be explained by pairing, which is not present in our theory, but an alternative theory involving only magnetic fluctuations has also been proposed \cite{Kharkov2019}.

\begin{figure}[tb]
\centering
\includegraphics[width=0.49\linewidth]{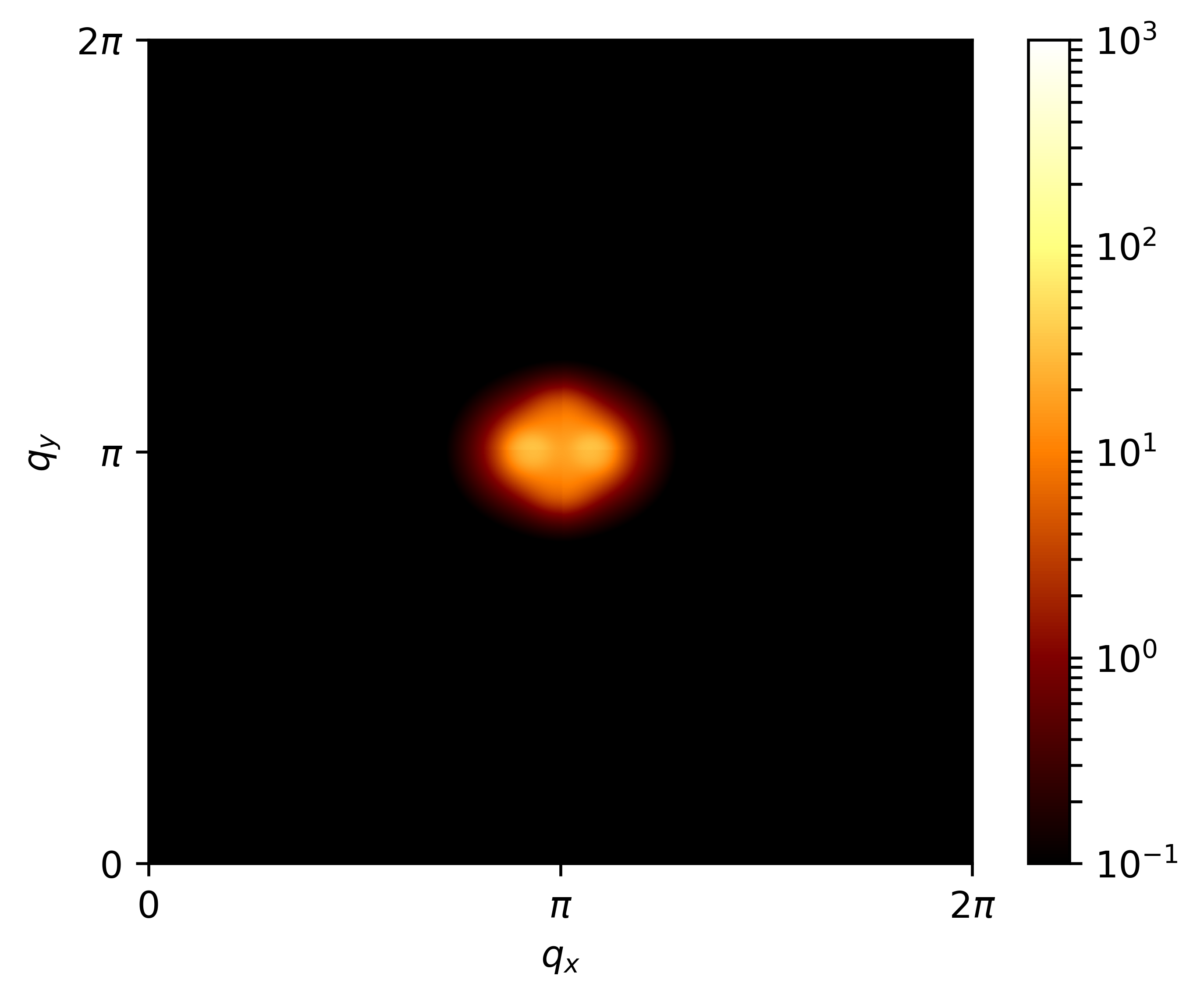}
\includegraphics[width=0.49\linewidth]{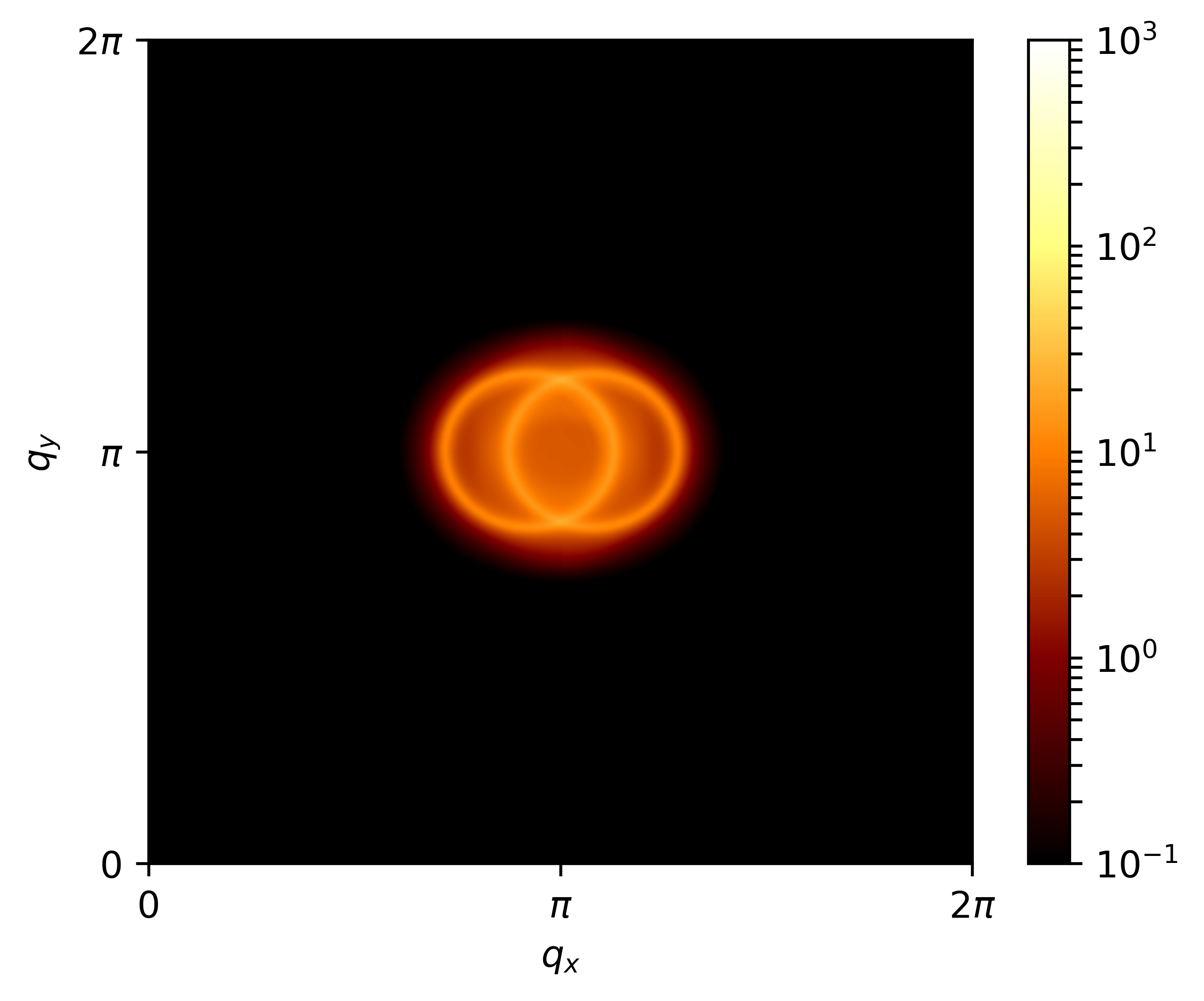}
 \caption{${\rm Im} S(\bq,\omega)$ as a function of momentum at two fixed frequencies $\omega = 0.10t$ (left) and $\omega = 0.20t$ (right) in the spiral state at $n = 0.95$ and $T = 0.02t$.
 The delta function contribution from the direct spinon term has again been broadened by a damping = 0.01.}
\label{fig: nematicity}
\end{figure}
While the SU(2) and translation symmetries are restored by the spinon fluctuations, nematic order, that is, a spontaneous breaking of the tetragonal symmetry of the square lattice, is inherited from the chargons if they undergo spiral order. In particular, ${\rm Im} S(\bq,\omega)$ exhibits a pronounced asymmetry between the $q_x$ and $q_y$ directions, as can be seen very clearly in Fig.~\ref{fig: nematicity}. Nematic order has indeed been observed in the spin structure factor of cuprates in the pseudogap regime \cite{Hinkov2004, Hinkov2008}.


\subsubsection{Static susceptibility}

At zero frequency, the spin susceptibility is real, and can be obtained directly (without analytic continuation) from Eq.~\eqref{eq: spin structure3} in the form
\begin{eqnarray} \label{eq: spin structure static}
 S(\bq,0) &=&
 \frac{1}{2} m^2 \left[ \cD(\bq+\bQ,0) + \cD(\bq-\bQ,0) \right] \nonumber \\
 &+& T \sum_{q'_0} \int_{\bq'}
 \cD(\bq',iq'_0) \, \tr\chi(\bq-\bq',-iq'_0) \, .
\end{eqnarray}

For $\bq \to 0$ this yields the static uniform spin susceptibility
$\kappa_s = \lim_{\bq \to 0} S(\bq,0)$ as
\begin{equation} \label{eq: kappa_s}
 \kappa_s = T \sum_{q_0} \int_\bq
 \cD(\bq,iq_0) \, \tr\chi(\bq,iq_0) \, .
\end{equation}
The static uniform spin susceptibility can be measured in experiments via the Knight shift.
Due to the momentum cutoff on the spinon propagators, the disconnected chargon susceptibility (the first line in Eq.~\eqref{eq: spin structure static}) does not contribute to $S(\bq,0)$ for small $\bq$, so that $S(\mathbf{0},0) = S_c(\mathbf{0},0)$.

\begin{figure}[tb]
\centering
 \includegraphics[width=0.49\linewidth]{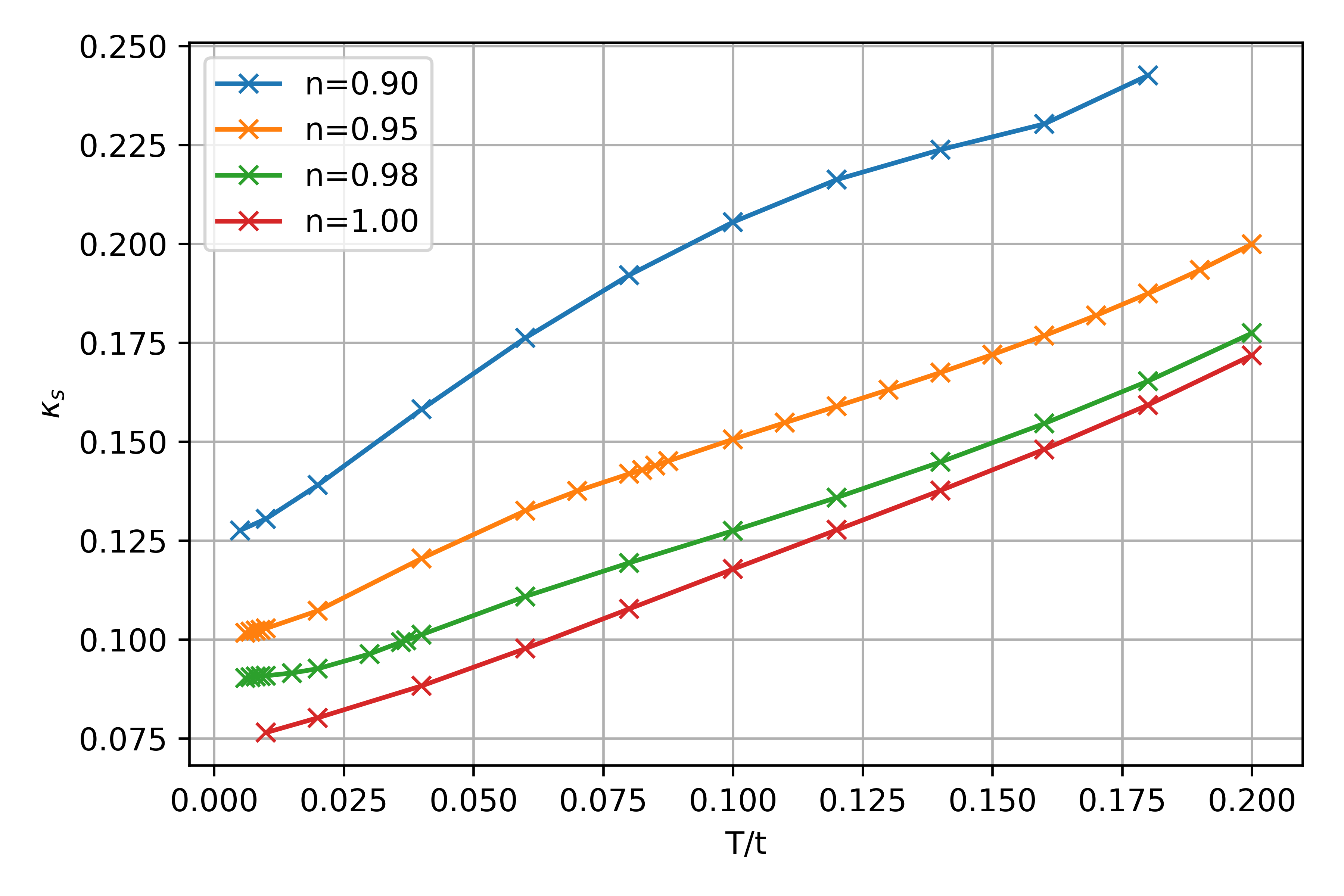}
 \caption{Uniform static spin susceptibility $\kappa_s$ as a function of temperature for various electron densities $n$.}
\label{fig: kappa_s}
\end{figure}
In Fig.~\ref{fig: kappa_s} we plot the uniform static spin susceptibility $\kappa_s$ of the electrons as a function of temperature for various densities. We see that $\kappa_s$ decreases monotonically upon lowering the temperature, as expected, and it saturates at a reduced but finite value for $T \to 0$. The electron spin susceptibility is generally smaller than the chargon spin susceptibility (see Fig.~\ref{fig: chi_u}), and it exhibits a stronger temperature dependence.

\begin{figure}[tb]
\centering
\includegraphics[width=0.49\linewidth]{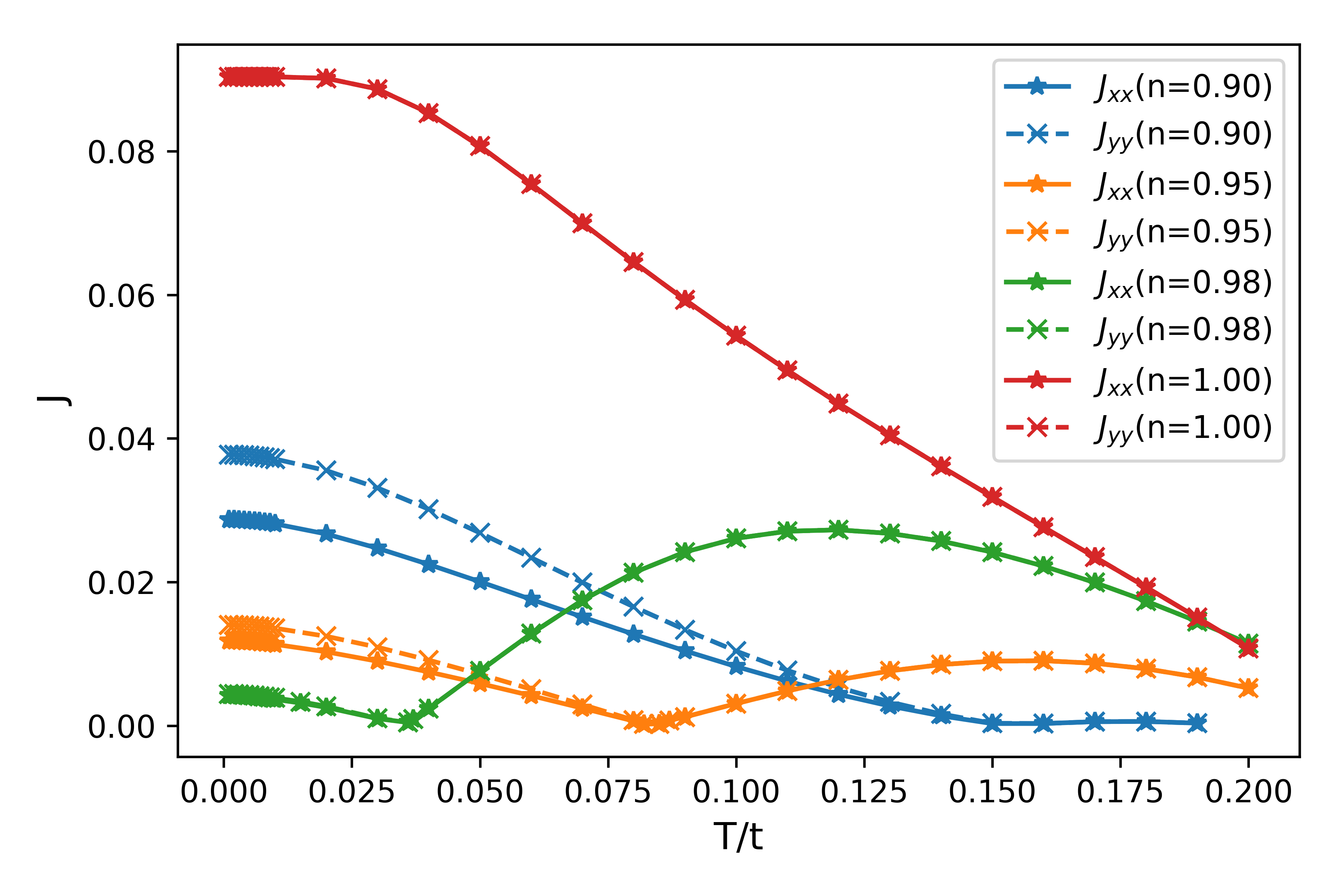}
\includegraphics[width=0.49\linewidth]{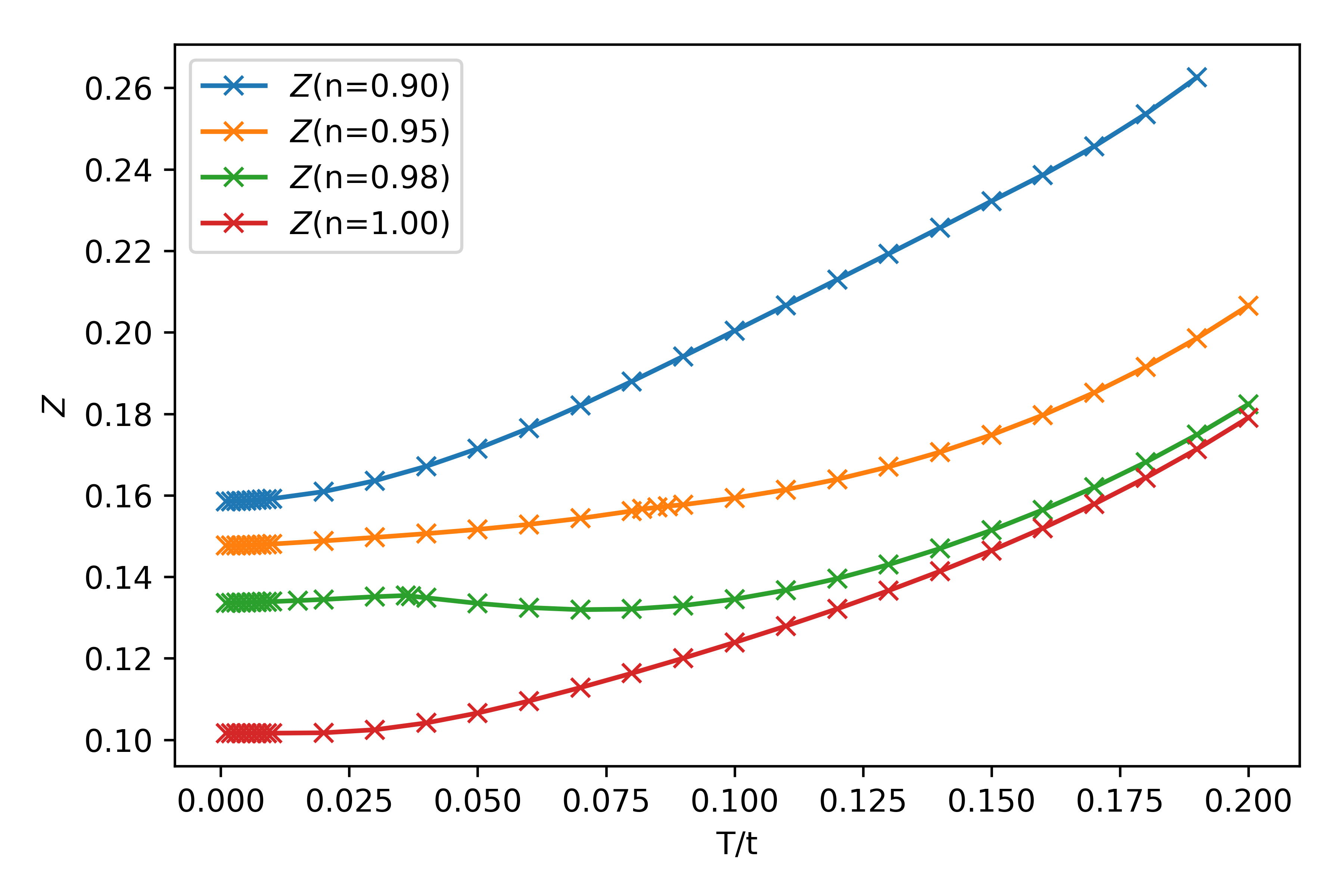}
 \caption{Spatial (left) and temporal (right) stiffnesses as functions of temperature for various electron densities. In the spiral regime, $J$ and $Z$ denote the spatial and temporal out-of-plane stiffnesses, respectively.}
\label{fig: stiffnesses}
\end{figure}
\begin{figure}[tb]
\centering
\includegraphics[width=0.49\linewidth]{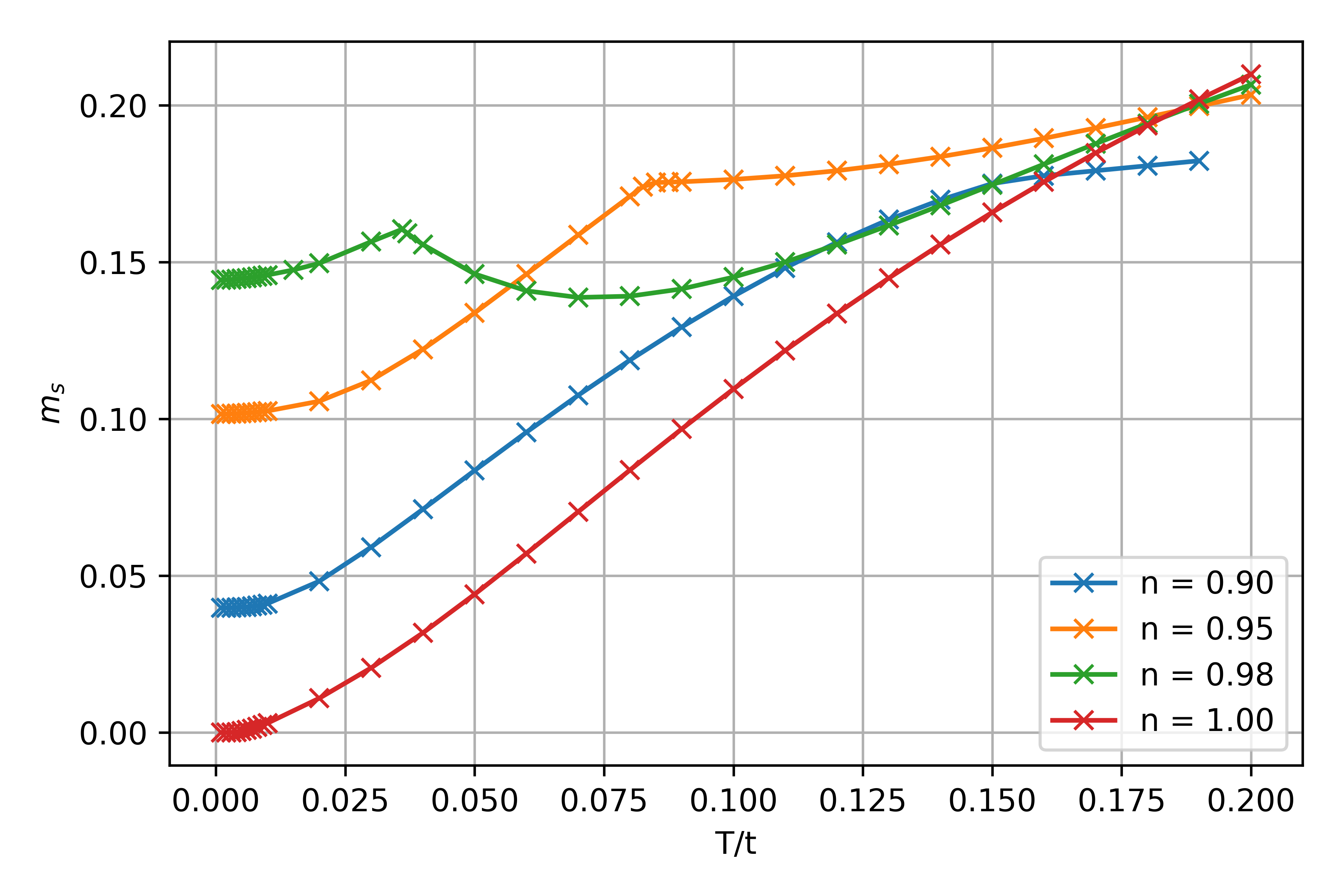}
 \caption{Spinon mass $m_s$ as a function of temperature for various electron densities and a fixed momentum cutoff $\Lam=0.3$.}
\label{fig: m_s}
\end{figure}
Since the temperature dependence of $\kappa_s$ is to a large extent determined by the spinon propagator $\cD(q)$ in Eq.~\eqref{eq: kappa_s}, it is instructive to compare to the temperature dependence of the quantities parametrizing $\cD(q)$. In Fig.~\ref{fig: stiffnesses} we plot the temperature dependence of the spatial and temporal stiffnesses, and in Fig.~\ref{fig: m_s} the temperature dependence of the spinon mass, which is determined by the sum rule $\int_q \cD(q) = \frac{1}{3}$. Only at half-filling the chargons order in a N\'eel state for all temperatures, while for the other electron densities a transition from N\'eel to spiral occurs at a critical temperature $T_{\rm nem}$ below $T^*$, see Fig.~\ref{fig: phasedia}. At this transition the spatial stiffness vanishes -- a generic feature which has been noticed and explained only quite recently \cite{Kharkov2018, Goremykin2024, Vilardi2025}. By contrast, the temperature stiffness $Z$ exhibits only a very mild kink at $T_{\rm nem}$. In the spiral regime we show only the out-of-plane stiffness $J^\perp$, since the in-plane stiffness does not affect the spinon propagator in our saddle-point approximation. The out-of-plane stiffness in the spiral regime is slightly anisotropic, that is, $J_{xx}^\perp \neq J_{yy}^\perp$.
For a fixed cutoff $\Lam$, the spinon mass $m_s$ is uniquely determined by the stiffnesses via the sum rule for the spinon propagator. A zero of $J$ at $T_{\rm nem}$ leads to an upward kink of $m_s$ at $T = T_{\rm nem}$. Depending on the ground state stiffnesses and the size of $\Lam$, the ground state can be ordered or quantum disordered. In the former case the spinon mass vanishes for $T \to 0$, while in the latter it remains finite.
For our choice of $\Lam$, the ground state is ordered and $m_s \to 0$ at half-filling, while $m_s$ remains finite for $T \to 0$ corresponding to a quantum disordered ground state for $n=0.9$, $0.95$, $0.98$. Note that the ground state spin gap is largest for the smallest doping value, because the ground state stiffnesses (both spatial and temporal) decrease with decreasing doping (see Fig.~\ref{fig: stiffnesses}). At half-filling the ground state spin gap jumps abruptly to zero because of a large discontinuity of the spatial stiffness \cite{Vilardi2025}.
In spite of the non-monotonic behavior of the spin stiffnesses and the spinon mass, both as a function of doping and temperature, the uniform static spin susceptibility decreases monotonically with temperature and doping.

Our result for $\kappa_s$ is consistent with measurements of the spin susceptibility in the pseudogap regime of cuprates, where a monotonic decrease to a reduced but finite value was observed upon lowering the temperature from $T^*$ to zero \cite{Zhou2025}.
However, on theoretical grounds our result is surprising, because a spin gap is expected to completely suppress the linear response to an external magnetic field at low temperatures.
Numerical simulations of the Hubbard model do not shed much light on this issue, because the low temperature limit cannot be accessed. For example, state-of-the-art cluster DMFT calculations reveal only a very mild decrease of $\kappa_s$ for temperatures $T < T^*$ in the lightly hole-doped regime, before the calculation has to be stopped at $T \approx 0.05t$ \cite{Chen2017}. A more pronounced decrease has been seen in a simulation using the recently developed minimally entangled typical termal states (METTS) method down to temperatures $T \approx 0.04t$ \cite{Wietek2021}.


\subsubsection{NMR relaxation rate}

The relaxation rate of nuclear spins due to the coupling to the electron spins can be obtained from the imaginary part of the spin susceptibility in the form
\begin{equation} \label{eq: NMR1}
 T_1^{-1} = \lim_{\omega \to 0} \frac{T}{\omega} \int_\bq
 F(\bq) \, {\rm Im} S(\bq,\omega) \, ,
\end{equation}
where $F(\bq)$ is a form factor related to the coupling between nuclear and electron spins. Note that we use natural units so that $\hbar = k_B = 1$.
We will see that the dominant contributions to $T_1^{-1}$ come from momenta near $\pm\bQ$, so that we can approximate $F(\bq)$ by a momentum independent number $F_Q$.

Due to the spinon mass, the first contribution to ${\rm Im} S(\bq,\omega)$ in Eq.~\eqref{eq: spin structure4}, originating from the disconnected chargon susceptibility, vanishes for $\omega \to 0$.
Inserting ${\rm Im} \, S_c(\bq,\omega)$ from Eq.~\eqref{eq: spin structure6}, we obtain
\begin{equation} \label{eq: NMR2}
 T_1^{-1} = \lim_{\omega \to 0} \frac{T}{2\omega} F_Q
 \sum_{s=\pm} \int_{\bq'}
 \frac{s}{Z \omega_{\bq'}}
 {\rm Im} \, \tr\chi_{\rm loc}(\omega-s\omega_{\bq'})
 \left[ b(s\omega_{\bq'}) - b(s\omega_{\bq'}-\omega) \right] \, ,
\end{equation}
with the trace of the local susceptibility
$\tr\chi_{\rm loc}(\omega) = \int_\bq \tr\chi(\bq,\omega)$. For $\bq \neq \pm \bQ$, the imaginary part of $\tr\chi(\bq,\omega)$ vanishes linearly for $\omega \to 0$. However, there is a larger contribution to ${\rm Im} \, \tr\chi_{\rm loc}(\omega)$ from the Goldstone poles at $\bq = \pm\bQ$. Here, we need to discuss the spiral and the N\'eel cases separately.

For $\bq$ near $\pm\bQ$ and low frequencies, the chargon spin susceptibilities in the spiral state exhibit the Goldstone poles \cite{Bonetti2022}
\begin{equation}
 \chi^a(\bq,\omega) \sim \frac{m^2/2}
 {J_{\alpha\beta}^a (q_\alpha \mp Q_\alpha)(q_\beta \mp Q_\beta) -
 Z^a (\omega + i0^+)^2} \, ,
\end{equation}
with $a \in \{\Box,\perp\}$,
where we have defined $\chi^\Box = \chi^{11} + \chi^{22}$ and $\chi^\perp = \chi^{33}$.
The imaginary parts of these poles are given by
\begin{equation}
 {\rm Im} \chi^a(\bq,\omega) \sim \frac{\pi}{4} \frac{m^2}{Z^a \Omega_{\bq\mp\bQ}^a}
 \left[ \delta(\omega - \Omega_{\bq\mp\bQ}^a) - \delta(\omega + \Omega_{\bq\mp\bQ}^a)
 \right] \, ,
\end{equation}
where
\begin{equation}
 \Omega_\bq^a = \sqrt{J_{\alpha\beta}^a q_\alpha q_\beta / Z^a} \, .
\end{equation}
Performing the elementary momentum integration, we obtain the contribution from the Goldstone modes to the imaginary part of the local chargon susceptibility
\begin{equation} \label{eq: chiloc}
 {\rm Im} \chi_{\rm loc}^a(\omega) = \frac{m^2}{4 \sqrt{\det J^a}} s(\omega) \, ,
\end{equation}
where $s(\omega)$ is the sign of $\omega$, and $J^a$ is the $2\times2$ matrix composed of $J_{\alpha\beta}^a$.
This Goldstone contribution remains finite for $\omega \to 0$, and thus dominates over all other contributions, which are linear in the frequency.

Inserting ${\rm Im} \chi_{\rm loc}^a(\omega)$ from Eq.~\eqref{eq: chiloc} into Eq.~\eqref{eq: NMR2}, and performing the sum over $s=\pm$, we obtain
\begin{equation} \label{eq: NMR3}
 T_1^{-1} = \lim_{\omega \to 0} \frac{T}{2\omega} F_Q
 \sum_{a \in \{\Box,\perp\}} \int_{\bq'}
 \frac{m^2}{4Z^a \sqrt{\det J^a} \, \omega_{\bq'}}
 \left[ b(\omega_{\bq'} - \omega) - b(\omega_{\bq'} + \omega) \right] \, .
\end{equation}
Using
\begin{equation}
 \lim_{\omega \to 0} \omega^{-1}
 \left[ b(\omega_{\bq'} - \omega) - b(\omega_{\bq'} + \omega) \right] =
 - 2 b'(\omega_{\bq'}) \, ,
\end{equation}
and the spinon density of states
\begin{equation}
 N_s(\omega) = \int_\bq \delta(\omega-\omega_\bq) =
 \frac{Z^\perp}{\sqrt{\det J^\perp}} \frac{\omega}{2\pi} \Theta(\omega - m_s) \, ,
\end{equation}
the momentum integral in Eq.~\eqref{eq: NMR3} can be carried out exactly, yielding the remarkably simple formula
\begin{equation} \label{eq: NMR4}
 T_1^{-1} = T F_Q \frac{m^2 \, b(m_s)}{4\pi \sqrt{\det J^\perp}}
 \sum_{a \in \{\Box,\perp\}} \frac{1}{\sqrt{\det J^a}} \, .
\end{equation}

In case of N\'eel order, there is no Goldstone pole in $\chi^{11}$, while $\chi^\perp = \chi^{22} + \chi^{33}$ has the form
\begin{equation}
 \chi^\perp(\bq,\omega) \sim \frac{2m^2}{J (\bq - \bQ)^2 - Z (\omega + i0^+)^2} \, ,
\end{equation}
with $\bQ = (\pi,\pi)$. The imaginary part of the local chargon susceptibility is then obtained from this single pole as
\begin{equation}
 {\rm Im} \, \tr\chi_{\rm loc}(\omega) = \frac{m^2}{2 \sqrt{\det J}} s(\omega) \, ,
\end{equation}
and the NMR relaxation rate becomes
\begin{equation} \label{eq: NMR5}
 T_1^{-1} = T F_Q \frac{m^2 \, b(m_s)}{2\pi \det J} \, .
\end{equation}
\begin{figure}[tb]
\centering
\includegraphics[width=0.5\linewidth]{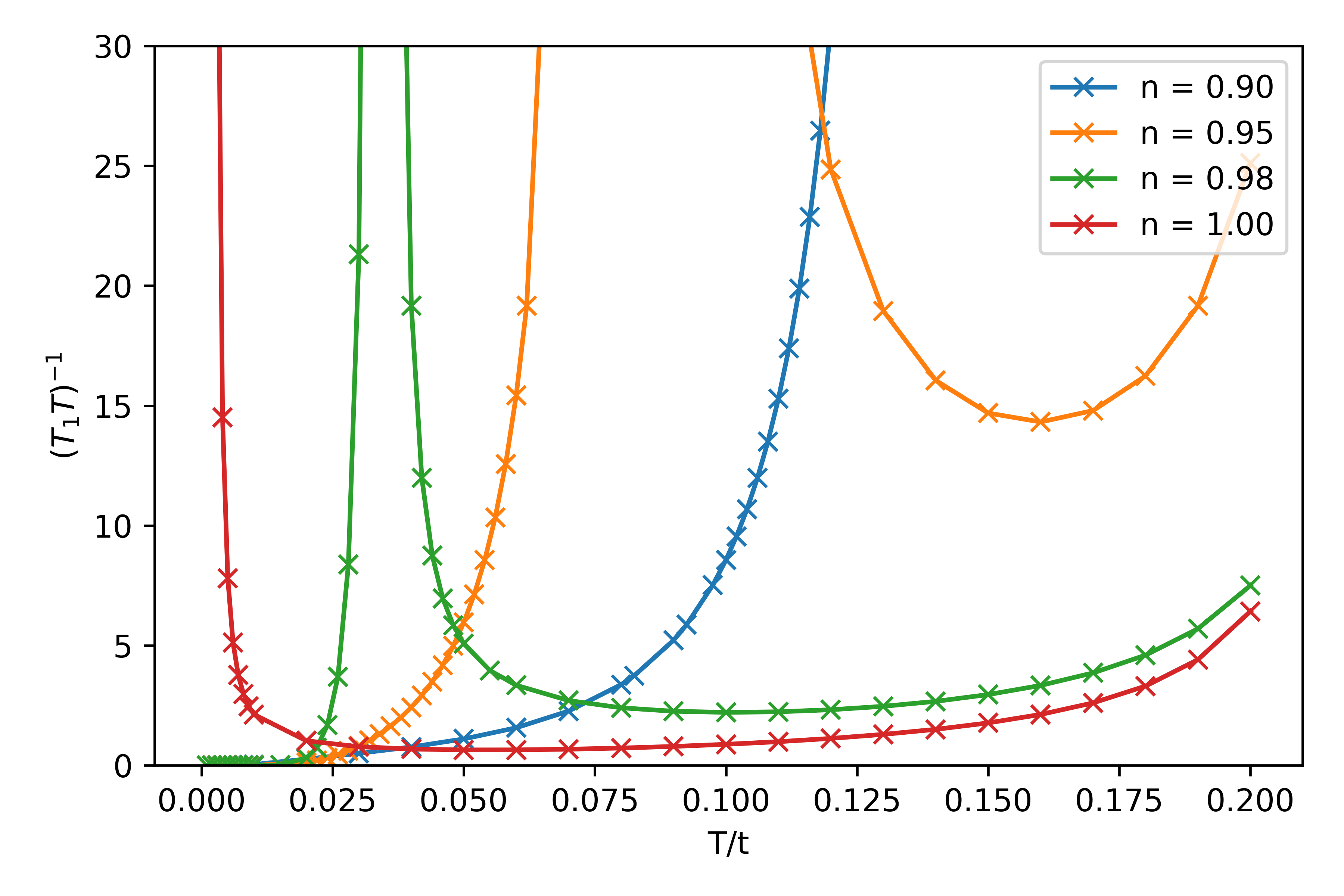}
 \caption{NMR relaxation rate $(TT_1)^{-1}$ as a function of temperature for various electron densities $n$. The form factor $F_Q$ has been set to one.}
\label{fig: T1}
\end{figure}
In Fig.~\ref{fig: T1} we show the NMR relaxation rate divided by $T$, that is,
$(TT_1)^{-1}$, as a function of temperature for various electron densities.
At half-filling, $(TT_1)^{-1}$ diverges for $T \to 0$ since the ground state is ordered so that the spinon mass $m_s$ vanishes exponentially. In layered systems such as the cuprates, this divergence is cut off by the coupling between the layers, which make the system three-dimensional at low energy scales.
In the doped cases, $(TT_1)^{-1}$ vanishes exponentially for $T \to 0$ due to the then finite spinon mass. The suppression of the NMR relaxation rate at low temperatures is one of the hallmarks of pseudogap behavior of cuprates \cite{Bankay1994}.
The divergences at finite temperatures are due to the vanishing spatial stiffnesses at the  N\'eel-to-spiral transition, see Fig.~\ref{fig: stiffnesses}. These divergences should be reduced to finite peaks by Landau damping and possibly other corrections to our approximations.


\section{Conclusions}

We have computed the electron spin susceptibility in the pseudogap regime of the two-dimensional Hubbard model at a moderate interaction strength. Approximate expressions for the dynamical spin susceptibility $S(\bq,\omega)$ have been derived in the framework of a recently formulated SU(2) gauge theory of fluctuating magnetic order \cite{Bonetti2022a}.
The gauge structure results from a fractionalization of the electron operators in fermionic chargons with a pseudospin degree of freedom and bosonic spinons \cite{Schulz1995, Dupuis2002, Borejsza2004, Sachdev2009}. The chargons are treated in a renormalized mean-field theory. They order in a N\'eel or spiral magnetic state in a broad doping range around half-filling below a doping-dependent transition temperature $T^*$. Fluctuations of the spin orientation are captured by a non-linear sigma model. The spin stiffnesses parametrizing the action of the sigma model are obtained from a renormalized RPA for the chargon susceptibility in the magnetic phases \cite{Bonetti2022}. The non-linear sigma model has been solved in a saddle point approximation.
Previous applications of the above \cite{Bonetti2022a} and related \cite{Sachdev2016, Chatterjee2017, Scheurer2018, Wu2018, Sachdev2019} gauge theories of fluctuating magnetic order have focused on the single-electron propagator, yielding a low-energy fermionic excitation spectrum concentrated on Fermi arcs.

The SU(2) spin symmetry is conserved by the spinon fluctuations at any finite temperature, in agreement with the Mermin-Wagner theorem. Whether the ground state is magnetically ordered or quantum disordered depends on the size on the stiffnesses, and on an ultraviolet cutoff required to regularize the non-linear sigma model. This cutoff remains a free parameter which we cannot compute within our theory.
Our results for the spin susceptibility share many features with experimental observations in the pseudogap regime of cuprates:
i) The dynamical spin susceptibility has a spin gap at finite temperatures, which suppresses ${\rm Im} S(\bq,\omega)$ at temperatures and energies below this gap,
ii) ${\rm Im} S(\bq,\omega)$ exhibits nematicity below a transition temperature $T_{\rm nem}$,
iii) The static uniform spin susceptibility $\kappa_s$ decreases substantially upon lowering the temperature below $T^*$, and
iv) the NMR relaxation rate $T_1^{-1}$ vanishes exponentially for $T \to 0$ if the ground state is quantum disordered.

At low hole doping the chargons undergo N\'eel order at $T^*$, which transforms into spiral order at a lower transition temperature $T_{\rm nem} < T^*$. By contrast, at larger hole doping, for our parameters above $p=0.1$ (see Fig.~\ref{fig: phasedia}), the chargons enter into a spiral state directly at $T^*$. Hence, in the former case the nematic transition occurs at a temperature below $T^*$, and in the former right at $T^*$. This dichotomy is in perfect agreement with experimental observations in cuprate compounds: A nematic transition right at $T^*$ has been detected in underdoped YBCO with hole concentrations between $p=0.11$ and $p=0.15$ \cite{Sato2017}, and also in moderately underdoped $\rm HgBa_2CuO_{4+\delta}$ \cite{Murayama2019}, while a nematic transition at a temperature $T_{\rm nem}$ well below $T^*$ has been found in strongly underdoped YBCO \cite{Hinkov2008}.

While the qualitative agreement of our results with the pseudogap behavior observed in experiments is encouraging, the finding of a finite residual uniform static spin susceptibility in the presence of a hard spin gap is suspicious. Indeed, the uniform susceptibility obtained from our approximations remains finite even at finite frequencies, violating an exact property following from the SU(2) spin symmetry \cite{Bonetti2022ward}. This implies that our theory is not conserving in the sense of Baym and Kadanoff \cite{Baym1961}. The source of this deficiency is probably the factorization of the electron spin susceptibility in chargon and spinon expectation values.

Our treatment of the chargons via a renormalized static mean-field theory is applicable only for weak or moderate interaction strengths. It is important to realize that pseudogap behavior does not require particularly strong interactions. An extension of our theory to strong coupling is possible by computing the chargon phase diagram and the corresponding spin stiffnesses from dynamical mean-field theory.


\section*{Acknowledgements}

We are grateful to Andrey Chubukov, Antoine Georges, Bernhard Keimer, Pavel Ostrovsky, Subir Sachdev, and Robin Scholle for valuable discussions. P.M.B. acknowledges support by the German National Academy of Sciences Leopoldina through Grant No. LPDS 2023-06, the Gordon and Betty Moore Foundation’s EPiQS Initiative Grant GBMF8683, and the U.S. National Science Foundation grant No. DMR-2245246.

\appendix


\section{Expressions for spin stiffnesses}
\label{app: spin stiffnesses}

The momentum derivatives on the right hand side of Eq.~\eqref{eq: Jin} for the spatial in-plane stiffness can be evaluated by applying them to the bare susceptibilities as given in Eq.~\eqref{eq: chi0t}. The Matsubara sums can be performed analytically.
For the derivative in the first term one obtains \cite{Bonetti2022ward}
\begin{equation} \label{eq: Jin0}
 \left. \partial^2_{q_\alpha q_\beta}
 \tilde{\chi}^{22}_0(\bq,0) \right|_{\bq=\mathbf{0}} =
 \int_\bk \gamma^\alpha_{\bk}\gamma^\beta_{\bk+\bQ} \!
 \left[ \frac{f(E_\bk^-) - f(E_\bk^+)}{8e^3_\bk} +
 \frac{f'(E_\bk^+) + f'(E_\bk^-)}{8e^2_\bk} \right] ,
\end{equation}
where $f'(x) = df/dx$, and $\gamma_\bk^\alpha = \partial_{k_\alpha} \epsilon_\bk$.
The momentum derivatives in the second term can be written as \cite{Bonetti2022ward}
\begin{eqnarray}
 \left. \partial_{q_\alpha} \tilde{\chi}_0^{20}(\bq,0) \right|_{\bq=\mathbf{0}} &=&
 - \left. \partial_{q_\alpha} \tilde{\chi}_0^{02}(\bq,0) \right|_{\bq=\mathbf{0}}
 \nonumber \\[1mm]
 &=& i \Delta_m \int_\bk \partial_{k_\alpha} h_\bk
 \left[ \frac{f(E_\bk^-) - f(E_\bk^+)}{8e^3_\bk} +
 \frac{f'(E_\bk^+) + f'(E_\bk^-)}{8e^2_\bk} \right] \, ,
\end{eqnarray}
and
\begin{eqnarray}
 \left. \partial_{q_\alpha} \tilde{\chi}_0^{21}(\bq,0) \right|_{\bq=\mathbf{0}} &=&
 - \left. \partial_{q_\alpha} \tilde{\chi}_0^{12}(\bq,0) \right|_{\bq=\mathbf{0}}
 \nonumber \\[2mm]
 &=& i \! \int_\bk h_\bk \partial_{k_\alpha} g_\bk \,
 \frac{f(E_\bk^-) - f(E_\bk^+)}{8e^3_\bk}
 + i \! \int_\bk \sum_\ell \left[
 h_\bk \partial_{k_\alpha} g_\bk + \ell e_\bk \partial_{k_\alpha} h_\bk \right]
 \frac{f'(E_\bk^\ell)}{8e_\bk^2} , \hskip 9mm
\end{eqnarray}
where $g_\bk = \frac{1}{2} \left( \xi_\bk + \xi_{\bk+\bQ} \right)$.

Applying the momentum derivatives to $\tilde{\chi}^{33}_0(\bq,0)$ in Eq.~\eqref{eq: Jout} for the spatial out-of-plane stiffness, and performing the Matsubara sum, yields \cite{Bonetti2022ward}
\begin{align}
 J^{\perp}_{\alpha\beta} &=
 \frac{1}{8}\int_\bk \sum_{\ell,\ell'=\pm} \left( 1 - \ell \frac{h_\bk}{e_\bk} \right)
 \left( 1 + \ell'\frac{h_{\bk+\bQ}}{e_{\bk+\bQ}} \right)
 \gamma^\alpha_{\bk+\bQ}\gamma^\beta_{\bk+\bQ}
 \frac{f(E^\ell_\bk) - f(E^{\ell'}_{\bk+\bQ})}{E^\ell_\bk - E^{\ell'}_{\bk+\bQ}}
 \nonumber \\
 & - \frac{1}{8}\int_\bk \sum_{\ell=\pm} \left[
 \left( 1 - \ell\frac{h_\bk}{e_\bk}\right)^2
 \gamma^\alpha_{\bk+\bQ} + \frac{\Delta_m^2}{e^2_\bk}\gamma^\alpha_\bk \right]
 \gamma^\beta_{\bk+\bQ}f'(E^\ell_\bk) \nonumber \\
 & - \frac{1}{8}\int_\bk \sum_{\ell=\pm} \left[\frac{\Delta_m^2}{e^2_\bk}
 (\gamma^\alpha_{\bk+\bQ}-\gamma^\alpha_\bk) \right] \gamma^\beta_{\bk+\bQ}
 \frac{f(E^\ell_\bk) - f(E^{-\ell}_\bk)}{E^\ell_\bk - E^{-\ell}_\bk} \, .
\end{align}

The frequency derivatives of bare susceptibility components on the right hand side of Eq.~\eqref{eq: Zin} for the temporal in-plane stiffness can be expressed as \cite{Bonetti2022ward}
\begin{eqnarray}
 2 \Delta_m^2 \left. \partial_\omega^2 \tilde\chi_0^{22}(\mathbf{0},\omega)
 \right|_{\omega=0} &=&
 2i \Delta_m \left. \partial_\omega \tilde\chi_0^{23}(\mathbf{0},\omega)
 \right|_{\omega=0} =
 \left. \tilde\chi_0^{33}(\mathbf{0},\omega) \right|_{\omega=0} \nonumber \\[1mm]
 &=& \Delta_m^2 \int_\bk \frac{f(E_\bk^-) - f(E_\bk^+)}{4e_\bk^3} \, .
\end{eqnarray}

Finally, the frequency derivatives of bare susceptibility components on the right hand side of Eq.~\eqref{eq: Zout} for the temporal out-of-plane stiffness can be written in the form \cite{Bonetti2022ward}
\begin{equation}
 \Delta_m^2 \left. \partial_\omega^2 \tilde\chi_0^{33}(\bQ,\omega) \right|_{\omega=0} =
 - \frac{1}{8} \int_\bk \sum_{\ell,\ell'=\pm}
 \left( 1 - \ell \frac{h_\bk}{e_\bk} \right)
 \left( 1 + \ell \frac{h_{\bk+\bQ}}{e_{\bk+\bQ}} \right)
 F_{\ell\ell'}(\bk,\bQ) \, ,
\end{equation}
where
\begin{equation}
 F_{\ell\ell'}(\bk,\bq) = {\rm Re} \, \frac{f(E_\bk^\ell) - f(E_{\bk+\bq}^{\ell'})}
 {E_\bk^\ell - E_{\bk+\bq}^{\ell'} + i0^+} \, ,
\end{equation}
and
\begin{eqnarray}
 \Delta_m \left. \partial_\omega \tilde\chi_0^{30}(\bQ,\omega) \right|_{\omega=0} &=&
 - \frac{1}{16} \int_\bk \sum_{\ell,\ell'=\pm}
 \left[ \ell \frac{\Delta_m}{e_\bk} + \ell' \frac{\Delta_m}{e_{\bk+\bQ}} +
 \ell\ell' \frac{\Delta_m (h_{\bk+\bQ} - h_\bk)}{e_\bk e_{\bk+\bQ}} \right]
 F_{\ell\ell'}(\bk,\bQ) \, , \nonumber \\
 \Delta_m \left. \partial_\omega \tilde\chi_0^{31}(\bQ,\omega) \right|_{\omega=0} &=&
 - \frac{1}{16} \int_\bk \sum_{\ell,\ell'=\pm}
 \left[ 1 + \ell \frac{h_\bk}{e_\bk} - \ell' \frac{h_{\bk+\bQ}}{e_{\bk+\bQ}} -
 \ell\ell' \frac{h_\bk h_{\bk+\bQ} - \Delta_m^2}{e_\bk e_{\bk+\bQ}} \right]
 F_{\ell\ell'}(\bk,\bQ) \, , \nonumber \\
 \Delta_m \left. \partial_\omega \tilde\chi_0^{32}(\bQ,\omega) \right|_{\omega=0} &=&
 \phantom{-} \frac{i}{16} \int_\bk \sum_{\ell,\ell'=\pm}
 \left[ 1 + \ell \frac{h_\bk}{e_\bk} - \ell' \frac{h_{\bk+\bQ}}{e_{\bk+\bQ}} -
 \ell\ell' \frac{h_\bk h_{\bk+\bQ} + \Delta_m^2}{e_\bk e_{\bk+\bQ}} \right]
 F_{\ell\ell'}(\bk,\bQ) \, . \nonumber \\
\end{eqnarray}
%


\section{\texorpdfstring{CP\/$^1$}{CP} representation of non-linear sigma model}
\label{app: CP1}

To prove general properties of the spinon propagator $D^{acbd}$ defined in Eq.~\eqref{eq: def D}, we make use of the CP$^1$ representation of the non-linear sigma model \cite{Auerbach1994}.


\subsection{\texorpdfstring{CP\/$^1$}{CP} representation of the action}

The matrix $\cR$ can be expressed as a triad of orthonormal unit vectors ${\bf\hat n}^a$, see Eq.~\eqref{eq: R triad}, which can be represented in terms of two complex Schwinger bosons $z_\up$ and $z_\down$ \cite{Sachdev1995},
\begin{subequations} \label{eq: ni}
\begin{align}
 & {\bf\hat n}^- = z (i\sigma^2\vec{\sigma}) z, \\
 & {\bf\hat n}^+ = z^* (i\sigma^2\vec{\sigma})^\dag z^*, \\
 & {\bf\hat n}^3 = z^* \vec{\sigma} z,
\end{align}
\end{subequations}
with $z = (z_\up,z_\down)$ and ${\bf\hat n}^\pm = {\bf\hat n}^1 \mp i {\bf\hat n}^2$.
The Schwinger bosons obey the constraint
\begin{equation} \label{eq: z boson constraint}
 z^*_\up z_\up + z^*_\down z_\down = 1 \, .
\end{equation}
The parametrization~\eqref{eq: ni} is equivalent to
\begin{equation} \label{eq: R to z}
 R = \left( \begin{array}{cc}
 z_\up &  -z_\down^* \\ z_\down & \phantom{-} z_\up^*
 \end{array} \right).
\end{equation}
Inserting the expressions \eqref{eq: R triad} and \eqref{eq: ni} into Eq.~\eqref{eq: sigma model} and assuming a stiffness matrix $\cJ_{\mu\nu}$ of the form~\eqref{eq: stiffness matrix}, we obtain the $\rm CP^1$ action for fluctuating spiral order
\begin{equation} \label{eq: CP1 action}
 \mathcal{S}_{\text{CP}^1}[z,z^*] = \int dx \,
 \Big[ 2J^\perp_{\mu\nu} [\partial_\mu z^*(x)] \partial_\nu z(x)
 - \, 2(J^\perp_{\mu\nu} - J^\Box_{\mu\nu}) j_\mu(x) j_\nu(x) \Big] \, ,
\end{equation}
with the current density
\begin{equation} \label{eq: current}
 j_\mu(x) = \frac{i}{2}
 \left[ z^*(x) \partial_\mu z(x) - z(x) \partial_\mu z^*(x) \right] \, .
\end{equation}
For the N\'eel case, the $\rm CP^1$ action is given by the same expression with
$J_{\mu\nu}^\perp = J_{\mu\nu}$ and $J_{\mu\nu}^\Box = 0$.

To see where the terms in Eq.~\eqref{eq: CP1 action} come from, we go through its derivation. Inserting $z = (z_\up,z_\down)$ into Eq.~\eqref{eq: ni} and carrying out the products yields
\begin{subequations}
\begin{align}
 \hat{\bf n}^- &= \left( z_\up z_\up - z_\down z_\down,
 i z_\up z_\up + i z_\down z_\down, - 2 z_\up z_\down \right) , \\
 \hat{\bf n}^+ &= \left( z_\up^* z_\up^* - z_\down^* z_\down^*,
 -i z_\up^* z_\up^* - i z_\down^* z_\down^*, - 2 z_\up^* z_\down^* \right) , \\
 \hat{\bf n}^3 &= \left( z_\up^* z_\down + z_\down^* z_\up,
 -i z_\up^* z_\down + i z_\down^* z_\up, z_\up^* z_\up - z_\down^* z_\down
 \right) .
\end{align}
\end{subequations}
Inserting this into the products of gradients in the non-linear sigma model yields
\begin{eqnarray}
 \partial_\mu \hat{\bf n}^3 \cdot \partial_\nu \hat{\bf n}^3 &=&
 2\partial_\mu (z_\up^* z_\down) \partial_\nu (z_\down^* z_\up) +
 \mu \lra \nu \nonumber \\ &+&
 \partial_\mu (z_\up^* z_\up - z_\down^* z_\down)
 \partial_\nu (z_\up^* z_\up - z_\down^* z_\down) \nonumber \\ &=&
 2 (z_\up \partial_\mu z_\up^*) (z_\down \partial_\nu z_\down^*) +
 2 (z_\down^* \partial_\mu z_\down) (z_\up^* \partial_\nu z_\up) + \mu \lra \nu
 \nonumber \\ &+&
 2|z_\down|^2 \partial_\mu z_\up^* \partial_\nu z_\up +
 2|z_\up|^2 \partial_\mu z_\down \partial_\nu z_\down^* + \mu \lra \nu
 \nonumber \\ &+&
 (z_\up \partial_\mu z_\up^*) (z_\up \partial_\nu z_\up^*) +
 (z_\up^* \partial_\mu z_\up) (z_\up^* \partial_\nu z_\up) + \up \, \lra \, \down
 \nonumber \\ &+&
 |z_\up|^2 \partial_\mu z_\up^* \partial_\nu z_\up +
 |z_\up|^2 \partial_\mu z_\up \partial_\nu z_\up^* + \up \, \lra \, \down
 \nonumber \\ &-&
 (z_\up \partial_\mu z_\up^* + z_\up^* \partial_\mu z_\up)
 (z_\down \partial_\nu z_\down^* + z_\down^* \partial_\nu z_\down) - \mu \lra \nu \, .
\end{eqnarray}
Using the constraint $|z_\up|^2 + |z_\down|^2 = 1$, this expression can be simplified to
\begin{eqnarray} \label{eq: dn3dn3}
 \partial_\mu \hat{\bf n}^3 \cdot \partial_\nu \hat{\bf n}^3 &=&
 2 \partial_\mu z_\up^* \partial_\nu z_\up +
 2 \partial_\mu z_\down \partial_\nu z_\down^* + \mu \lra \nu
 \nonumber \\ &-&
 |z_\up|^2 \partial_\mu z_\up^* \partial_\nu z_\up -
 |z_\up|^2 \partial_\mu z_\up \partial_\nu z_\up^* - \up \, \lra \, \down
 \nonumber \\ &+&
 (z_\up \partial_\mu z_\up^*) (z_\down \partial_\nu z_\down^*) +
 (z_\down^* \partial_\mu z_\down) (z_\up^* \partial_\nu z_\up) + \mu \lra \nu
 \nonumber \\ &+&
 (z_\up \partial_\mu z_\up^*) (z_\up \partial_\nu z_\up^*) +
 (z_\up^* \partial_\mu z_\up) (z_\up^* \partial_\nu z_\up) + \up \, \lra \, \down
 \nonumber \\ &-&
 (z_\up \partial_\mu z_\up^*) (z_\down^* \partial_\nu z_\down) -
 (z_\up^* \partial_\mu z_\up) (z_\down \partial_\nu z_\down^*) - \mu \lra \nu
 \nonumber \\ &=&
 2 \partial_\mu z^* \partial_\nu z + 2 \partial_\nu z^* \partial_\mu z -
 4 j_\mu j_\nu \, .
\end{eqnarray}
Similarly, for $\partial_\mu \hat{\bf n}^+ \cdot \partial_\nu \hat{\bf n}^-$ we obtain
\begin{eqnarray}
 \partial_\mu \hat{\bf n}^+ \cdot \partial_\nu \hat{\bf n}^- &=&
 2 \partial_\mu (z_\up^* z_\up^*) \partial_\nu (z_\up z_\up) +
 2 \partial_\mu (z_\down^* z_\down^*) \partial_\nu (z_\down z_\down)
 \nonumber \\ &+&
 4 \partial_\mu (z_\up^* z_\down^*) \partial_\nu (z_\up z_\down)
 \nonumber \\ &=&
 8 (z_\up \partial_\mu z_\up^*) (z_\up^* \partial_\nu z_\up) + \up \lra \down
 \nonumber \\ &+&
 4 (z_\up \partial_\mu z_\up^*) (z_\down^* \partial_\nu z_\down) +
 4 |z_\down|^2 \partial_\mu z_\up^* \partial_\nu z_\up + \up \lra \down
 \nonumber \\ &=&
 4 \partial_\mu z_\up^* \partial_\nu z_\up + \up \lra \down \nonumber \\ &+&
 4 (z_\up \partial_\mu z_\up^*) (z_\up^* \partial_\nu z_\up) +
 4 (z_\up \partial_\mu z_\up^*) (z_\down^* \partial_\nu z_\down) + \up \lra \down \, .
\end{eqnarray}
The constraint $z^*z = 1$ implies $z \partial_\mu z^* = - z^* \partial_\mu z$, so that the current from Eq.~\eqref{eq: current} can be rewritten in the form
$j_{\mu} = - i z \partial_\mu z^*$ or $j_{\mu} = i z^* \partial_\mu z$.
Using these relations we find
\begin{equation}
 \partial_\mu \hat{\bf n}^+ \cdot \partial_\nu \hat{\bf n}^- =
 4 \partial_\mu z^* \partial_\nu z + 4 j_\mu j_\nu \, ,
\end{equation}
and thus
\begin{eqnarray} \label{eq: dn1dn1}
 \partial_\mu \hat{\bf n}^1 \cdot \partial_\nu \hat{\bf n}^1 +
 \partial_\mu \hat{\bf n}^2 \cdot \partial_\nu \hat{\bf n}^2 &=&
 \frac{1}{2} \left(
 \partial_\mu \hat{\bf n}^+ \cdot \partial_\nu \hat{\bf n}^- +
 \partial_\mu \hat{\bf n}^- \cdot \partial_\nu \hat{\bf n}^+ \right)
 \nonumber \\ &=&
 2 \partial_\mu z^* \partial_\nu z + 2 \partial_\nu z^* \partial_\mu z +
 4 j_\mu j_\nu \, .
\end{eqnarray}
Inserting Eqs.~\eqref{eq: dn3dn3} and \eqref{eq: dn1dn1} into the non-linear sigma model \eqref{eq: S[R]} with the coefficients $\cP_{\mu\nu}^{aa}$ for spiral order from Eq.~\eqref{eq: P_spiral}, one obtains the CP$^1$ representation of the action, Eq.~\eqref{eq: CP1 action}.
For N\'eel order aligned in $z$ direction, one obtains Eq.~\eqref{eq: CP1 action} with $J^\perp = J$ and $J^\Box = 0$. To obtain the same CP$^1$ action for N\'eel order in the $x$ direction, as assumed in the main text, one needs to permute the assignment in Eq.~\eqref{eq: ni}, so that ${\bf\hat n}^1 = z^* \vec{\sigma} z$ and
${\bf\hat n}^\pm = {\bf\hat n}^2 \mp i {\bf\hat n}^3$.


\subsection{Spinon correlator}

To evaluate the spinon correlator $\bra \cR_j^{ac}(\tau) \cR_{j'}^{bd}(0) \ket$ in Eq.~\eqref{eq: suscept fac}, we now use the CP$^1$ representation of the rotation matrices, Eqs.~\eqref{eq: R triad} and \eqref{eq: ni}. The spinon correlator thus becomes a 4-point function of Schwinger bosons.
We consider only the case without a spinon condensate, which applies at any non-zero temperature, and also to a quantum disordered ground state. In this case only ``normal'' expectation values with an equal number of $z$ and $z^*$ bosons contribute, that is, only
$\bra \hat{\bf n}_j^\pm(\tau) \hat{\bf n}_{j'}^\mp(0) \ket$ and
$\bra \hat{\bf n}_j^3(\tau) \hat{\bf n}_{j'}^3(0) \ket$ need to be computed.

The SU(2) symmetry imposes the following general spin structure on normal boson 4-point functions:
\begin{equation} \label{eq: vertex symmetry}
 \bra z_{s_{1'}}^*(1') z_{s_1}^{\phantom *}(1)
 z_{s_{2'}}^*(2') z_{s_2}^{\phantom *}(2) \ket =
 \Gamma(1'2';12) \delta_{s_1s_{1'}} \delta_{s_2s_{2'}} +
 \Gamma(2'1';12) \delta_{s_1s_{2'}} \delta_{s_2s_{1'}} \, ,
\end{equation}
where $1,2,1',2'$ is a shorthand for the space and time variables.
This symmetry relation is the bosonic analogue of a well-known property of the fermion 4-point vertex in SU(2) symmetric Fermi systems \cite{Salmhofer2001, Eberlein2010}.
Using this relation, the contributing expectation values can be written as
\begin{eqnarray}
 \bra n_a^+(1) n_b^-(2) \ket &=&
 \bra z_{s_{1'}}^*(1) z_{s_1}^*(1)
 z_{s_{2'}}^{\phantom *}(2) z_{s_2}^{\phantom *}(2) \ket
 (i\sg^2\sg^a)_{s_{1'}s_1}^\dag (i\sg^2\sg^b)_{s_{2'}s_2}^{\phantom\dag} \nonumber \\
 &=& \Gamma(11;22) \left(
 \delta_{s_1s_2} \delta_{s_{1'}s_{2'}} + \delta_{s_1s_{2'}} \delta_{s_{1'}s_2} \right)
 (\sg^a\sg^2)_{s_{1'}s_1} (\sg^2\sg^b)_{s_{2'}s_2}  \nonumber \\
 &=& 4 \Gamma(11;22) \delta_{ab} \, ,
\end{eqnarray}
and, analogously, $\bra n_a^-(1) n_b^+(2) \ket = 4 \Gamma(22;11) \delta_{ab}$, such that
\begin{subequations}
\begin{align}
 & \bra n_a^1(1) n_b^1(2) \ket = \bra n_a^2(1) n_b^2(2) \ket =
 \left[ \Gamma(11;22) + \Gamma(22;11) \right] \delta_{ab} \, , \\
 & \bra n_2^a(1) n_1^b(2) \ket = - \bra n_1^a(1) n_2^b(2) \ket =
 i \left[ \Gamma(11;22) - \Gamma(22;11) \right] \delta_{ab} = 0 \, .
\end{align}
\end{subequations}
The last term vanishes since $\Gamma(11;22) = \Gamma(22;11)$ due to time reversal and spatial reflection symmetry.
For the third component we obtain
\begin{eqnarray}
 \bra n_a^3(1) n_b^3(2) \ket &=&
 \bra z_{s_{1'}}^*(1) z_{s_1}^{\phantom *}(1)
 z_{s_{2'}}^*(2) z_{s_2}^{\phantom *}(2) \ket
 \sg_{s_{1'}s_1}^a \sg_{s_{2'}s_2}^b  \nonumber \\
 &=& \left[ \Gamma(12;12) \delta_{s_1s_{1'}} \delta_{s_2s_{2'}} +
 \Gamma(21;12) \delta_{s_1s_{2'}} \delta_{s_2s_{1'}} \right]
 \sg_{s_{1'}s_1}^a \sg_{s_{2'}s_2}^b \nonumber \\
 &=& \Gamma(12;12) \tr(\sg^a) \tr(\sg^b) + \Gamma(21;12) \tr(\sg^a \sg^b)
 = 2 \Gamma(21;12) \delta_{ab} \, .
\end{eqnarray}

The relations derived above can be summarized as
\begin{equation}
 \cD^{acbd}(1,2) = \bra n_a^c(1) n_b^d(2) \ket =
 \delta_{ab} \delta_{cd} \, \cD^c(1,2) \, ,
\end{equation}
with
\begin{subequations}
\begin{align}
 & \cD^1(1,2) = \cD^2(1,2) = 2 \Gam(11;22) \, , \\
 & \cD^3(1,2) = 2 \Gam(21;12) \, .
\end{align}
\end{subequations}

Due to the symmetry relation Eq.~\eqref{eq: vertex symmetry}, we can express $\Gamma(11;22)$ and $\Gamma(21;12)$ by various, at first sight distinct expectation values.
Inserting $1 = (j,\tau)$ and $2 = (j',0)$, we choose
\begin{subequations} \label{eq: spinon corr}
\begin{align}
 & \Gamma(11;22) = \bra z_{j\up}^*(\tau) z_{j\down}^*(\tau)
 z_{j'\down}^{\phantom *}(0) z_{j'\up}^{\phantom *}(0) \ket
 \equiv \Gamma_{jj'}(\tau)  \, , \\
 & \Gamma(21;12) = \bra z_{j\up}^*(\tau) z_{j\down}^{\phantom *}(\tau)
 z_{j'\down}^*(0) z_{j'\up}^{\phantom *}(0) \ket
 \equiv \Gamma'_{jj'}(\tau)  \, .
\end{align}
\end{subequations}
An advantage of this choice is that in a perturbative expansion of these expectation values no disconnected Feynman diagrams contribute, since the spinon propagators are spin conserving.

Inserting the above results for the spinon correlator into the factorized susceptibility, Eq.~\eqref{eq: suscept fac}, we obtain
\begin{equation} \label{eq: spin structure1a}
 S_{jj'}^{ab}(\tau) =
 2 \Gamma_{jj'}(\tau) \left[ \bar\chi_{jj'}^{11}(\tau) + \bar\chi_{jj'}^{22}(\tau) \right]
 \delta_{ab} +
 2 \Gamma'_{jj'}(\tau) \bar\chi_{jj'}^{33}(\tau) \delta_{ab} \, .
\end{equation}
This result is manifestly spin rotation invariant. The SU(2) symmetry of the spinons thus restores the broken symmetry of the chargon susceptibility.
Inserting the disconnected contributions to $\bar\chi^{aa}(\tau)$ as in Eq.~\eqref{eq: chi bar chi}, we find, for a spiral state aligned in the $xy$ plane,
\begin{equation} \label{eq: spin structure2a}
 S_{jj'}^{ab}(\tau) =
 2 \Gamma_{jj'}(\tau) \left[ \chi_{jj'}^{11}(\tau) + \chi_{jj'}^{22}(\tau) +
 m^2 \cos(\bQ (\br_j - \br_{j'})) \right]
 \delta_{ab} +
 2 \Gamma'_{jj'}(\tau) \chi_{jj'}^{33}(\tau) \delta_{ab} \, .
\end{equation}
$S_{jj'}^{ab}(\tau)$ is translation invariant, unlike $\bar\chi_{jj'}^{11}$ and $\bar\chi_{jj'}^{22}$.
For the disconnected contributions this is obvious from Eq.~\eqref{eq: spin structure2a}.
For the connected parts, the translation invariance follows from the translation invariance of the chargon susceptibility $\tilde\chi_{jj'}$ in the rotated spin basis, in combination with $\chi_{jj'}^{33} = \tilde\chi_{jj'}^{33}$ and the invariance of the trace of the susceptibility under a change of basis.

It is instructive to see what happens if we were to average only over \emph{global}, space and time independent, rotations in Eq.~\eqref{eq: suscept fac}. In this case the spinon correlators in Eq.~\eqref{eq: spinon corr} can be calculated exactly.
One can show directly via a complex integration over $z_\up$ and $z_\down$ that $\Gamma_{jj'}(\tau) = \Gamma'_{jj'}(\tau) = \frac{1}{6}$ for global rotations, so that
$S_{jj'}^{ab}(\tau) = \frac{1}{3} \tr\left[\bar\chi_{jj'}(\tau)\right] \delta_{ab}$,
which is the intuitively expected result.

The quadratic part of the CP$^1$ action \eqref{eq: CP1 action} is symmetric under $z_\down \lra z_\down^*$. Moreover, the transformation $z_\down \lra z_\down^*$ transforms $\Gam_{jj'}(\tau)$ into $\Gam'_{jj'}(\tau)$, and vice versa. Hence, neglecting the current-current interaction in Eq.~\eqref{eq: CP1 action}, one obtains $\Gam_{jj'}(\tau) = \Gam'_{jj'}(\tau)$.


\section{Saddle point solution in \texorpdfstring{CP\/$^1$}{CP} representation}
\label{app: CP1 saddle}

As an alternative to the saddle point solution of the non-linear sigma model in the adjoint representation, as used in the main text, here we discuss another saddle point solution, which is based on the {\em CP}\/$^1$ representation of the model \cite{Auerbach1994, Chubukov1994}.


\subsection{Large {\em N}\/ limit and saddle point}

The current-current interaction in Eq.~\eqref{eq: CP1 action} can be decoupled by a Hubbard-Stratonovich transformation with a U(1) gauge field $\mathcal{A}_\mu$, and the constraint \eqref{eq: z boson constraint} can be implemented by a Lagrange multiplier field $\lambda$. The resulting action describes the massive $\rm CP^1$ model~\cite{Azaria1995}
\begin{eqnarray} \label{eq: massive CP1 model}
 \cS_{\text{CP}^1}[z,z^*,\mathcal{A}_\mu,\lambda] &=& \int dx
 \Big[ 2J^\perp_{\mu\nu} (D_\mu z)^* (D_\nu z) \nonumber \\
 &+& \frac{1}{2} M_{\mu\nu} \mathcal{A}_\mu \mathcal{A}_\nu + i\lambda(z^*z-1)
 \Big] \, , \hskip 5mm
\end{eqnarray}
where $D_\mu = \partial_\mu - i\mathcal{A}_\mu$ is the covariant derivative. We have suppressed the space-time dependence of the fields to shorten the expression. The coefficients $M_{\mu\nu}$ are the matrix elements of the mass tensor of the U(1) gauge field,
\begin{equation}
 {\rm M} = 4 \left[ ({\rm J}^\Box)^{-1} - ({\rm J}^\perp)^{-1} \right]^{-1} \, ,
\end{equation}
where ${\rm J}^\Box$ and ${\rm J}^\perp$ are the stiffness tensors built from the matrix elements
$J_{\mu\nu}^\Box$ and $J_{\mu\nu}^\perp$, respectively.
In case of N\'eel order this mass tensor vanishes and the CP$^1$ action is U(1) gauge invariant.

To perform a large $N$ expansion, we extend the two-component field $z = (z_\up,z_\down)$ to an $N$-component field $z = (z_1,\dots,z_N)$, and rescale it by a factor $\sqrt{N/2}$ so that it now satisfies the local constraint
\begin{equation} \label{eq: constraint}
 z^*(x) z(x) = \sum_{\sg=1}^N z^*_\sg(x) z_\sg(x) = \frac{N}{2} \, .
\end{equation}
To obtain a nontrivial limit $N \to \infty$, we rescale also the stiffnesses $J^\perp_{\mu\nu}$ and $J^\Box_{\mu\nu}$ by a factor $2/N$, yielding the massive CP$^{N-1}$ action
\begin{eqnarray} \label{eq: massive CPN1 model}
 \cS_{\text{CP}^{N-1}}[z,z^*,\mathcal{A}_\mu,\lambda] &=& \int dx
 \Big[ 2J_{\mu\nu}^\perp (D_\mu z)^* (D_\nu z) \nonumber \\
 &+& \frac{N}{4} M_{\mu\nu} \mathcal{A}_\mu \mathcal{A}_\nu +
 i\lambda \left( z^*z - N/2 \right)
 \Big] . \hskip 7mm
\end{eqnarray}

In the large $N$ limit, the partition function of the CP$^{N-1}$ model is dominated by a saddle point for the fields $\mathcal{A}$ and $\lambda$, at which $\mathcal{A}_\mu(x) = 0$, and $\lambda(x) = \lam$ is uniform in space and time \cite{Auerbach1994, Azaria1995}.
One can then read off the spinon propagator directly from the action \eqref{eq: massive CPN1 model},
\begin{equation} \label{eq: spinon propag}
 D(q) = \bra z_\sg^*(q) z_\sg(q) \ket = \frac{1}{2Z^\perp q_0^2 +
 2J_{\alpha\beta}^\perp q_\alpha q_\beta + 2 Z^\perp m_s^2} \, ,
\end{equation}
where $2m_s^2 = i\lam/Z^\perp$.
The Lagrange multiplier $\lam$ and thus the spinon mass $m_s$ are fixed by taking the average of the constraint, leading to $\bra z^*(x) z(x) \ket = N/2$, and thus
$\int_q D(q) = \frac{1}{2}$.
This relation for the spinon mass has the same structure as the corresponding condition
$\int_q \cD(q) = \frac{1}{3}$ with $\cD(q)$ from Eq.~\eqref{eq: cD spiral}, but the right hand side is larger leading to a somewhat smaller spinon mass.


\subsection{Spinon correlator in large {\em N}\/ limit}

We still need to evaluate the correlation functions $\Gamma_{jj'}(\tau)$ and $\Gamma'_{jj'}(\tau)$ defined in Eq.~\eqref{eq: spinon corr}. In accordance with our approximate solution of the non-linear sigma model, we will do this by employing the CP$^{N-1}$ action, Eq.~\eqref{eq: massive CPN1 model}, in the large $N$ limit.

The CP$^{N-1}$ action contains interactions between the Schwinger bosons and the fields $\cA_\mu$ and $\lam$. However, there are no contributions involving these interactions to the correlators $\Gamma_{jj'}(\tau)$ and $\Gamma'_{jj'}(\tau)$ in the large $N$ limit. This can be easily seen by considering the contributing Feynman diagrams. An important point here is that the interaction vertices conserve the spin quantum number of the Schwinger bosons.
Hence, the 4-point correlators in Eq.~\eqref{eq: spinon corr} can be simply factorized in the large $N$ limit, yielding
\begin{equation}
 \Gam_{jj'}(\tau) = \Gam'_{jj'}(\tau) = \left[ D_{jj'}(\tau) \right]^2 \, ,
\end{equation}
where $D_{jj'}(\tau) = \bra z_{j\sg}^*(\tau) z_{j'\sg}^{\phantom *}(0) \ket$, and we have used the inversion symmetries in space and time $D_{jj'}(\tau) = D_{j'j}(\tau)$ and $D_{jj'}(\tau) = D_{jj'}(-\tau)$.
Eq.~\eqref{eq: spin structure2a} can thus be simplified to
\begin{equation} \label{eq: spin structure3a}
 S_{jj'}^{ab}(\tau) = 2 \left[ D_{jj'}(\tau) \right]^2 \left[
 m^2 \cos(\bQ(\br_j - \br_{j'})) + \tr(\chi_{jj'}(\tau)) \right] \delta_{ab} \, .
\end{equation}
This expression has the same form as Eq.~\eqref{eq: spin structure2}, with the spinon propagator in the adjoint representation $\cD_{jj'}(\tau)$ now replaced by
$2[D_{jj'}(\tau)]^2$, where $D_{jj'}(\tau)$ is the spinon propagator in the $\rm CP^1$ representation of the SU(2) group.

Ultimately we are interested in the Fourier transform $S^{ab}(\bq,iq_0)$ of the spin susceptibility $S_{jj'}^{ab}(\tau)$, and its analytic continuation to real frequencies $S^{ab}(\bq,\omega)$. The product of spinon and chargon correlators in Eq.~\eqref{eq: spin structure3a} becomes a convolution in Fourier space. In the remainder of this section we therefore evaluate the Fourier transform of
$\Pi_{jj'}^s(\tau) = \left[ D_{jj'}(\tau) \right]^2$, which is also given by a convolution,
\begin{equation}
 \Pi^s(\bq,iq_0) = T \sum_{q'_0} \int_{\bq'}
 D(\bq',iq'_0) D(\bq'+\bq,iq'_0 + iq_0) \, ,
\end{equation}
with the spinon propagator from Eq.~\eqref{eq: spinon propag}.
Carrying out the Matsubara sum, and continuing analytically to real frequencies, one obtains
\begin{equation} \label{eq: Pis}
 \Pi^s(\bq,\omega) = \int_{\bq'} \sum_{\ell,\ell'=\pm}
 \frac{\ell\ell'}{(2Z^\perp)^2 \, \omega_{\bq'}^s \omega_{\bq'+\bq}^s} \,
 \frac{b(\ell\omega_{\bq'+\bq}^s) - b(\ell'\omega_{\bq'}^s)}
 {\omega + i0^+ + \ell\omega_{\bq'+\bq}^s - \ell'\omega_{\bq'}^s} \, ,
\end{equation}
where $b(x) = \left(e^{x/T} - 1 \right)^{-1}$ is the Bose-Einstein distribution, and
\begin{equation}
 \omega_\bq^s = \sqrt{(J_{\alpha\beta}^\perp/Z^\perp) q_\alpha q_\beta + m_s^2}
\end{equation}
is the spinon dispersion.

Fourier transforming Eq.~\eqref{eq: spin structure3a} yields
$S^{ab}(q) = S(q) \delta_{ab}$ with
\begin{eqnarray} \label{eq: spin structure4a}
 S(\bq,iq_0) &=&
 m^2 \left[ \Pi^s(\bq+\bQ,iq_0) + \Pi^s(\bq-\bQ,iq_0) \right] \nonumber \\
 &+& 2 \sum_{q'_0} \int_{\bq'}
 \Pi^s(\bq',iq'_0) \, \tr\chi(\bq-\bq',iq_0-iq'_0) \, ,
\end{eqnarray}
and its analytic continuation to real frequencies
\begin{eqnarray} \label{eq: spin structure5a}
 {\rm Im} S(\bq,\omega) &=&
 m^2 \, {\rm Im} \left[ \Pi^s(\bq+\bQ,\omega) + \Pi^s(\bq-\bQ,\omega) \right]
 \nonumber \\
 &+& \frac{2}{\pi} \int d\omega' \int_{\bq'}
 {\rm Im} \Pi^s(\bq',\omega') \, {\rm Im} \, \tr\chi(\bq-\bq',\omega-\omega')
 \left[ b(\omega') - b(\omega'-\omega) \right] . \hskip 5mm
\end{eqnarray}
These relations have the same form as the corresponding relations \eqref{eq: spin structure3} and \eqref{eq: spin structure4} in the main text, with $\cD(q)$ replaced by $2\Pi^s(q)$.

For $T \to 0$ the Bose functions in Eq.~\eqref{eq: Pis} become step functions,
$b(x) \to - \Theta(-x)$, so that only terms with $\ell' = - \ell$ contribute to $\Pi^s(\bq,\omega)$. Hence, the imaginary part ${\rm Im} \Pi^s(\bq,\omega)$ vanishes for $|\omega| < 2m_s$ in the low temperature limit, and this spin gap carries over to ${\rm Im} S(\bq,\omega)$. In this respect the result for the electron spin susceptibility obtained from the saddle point in the $\rm CP^1$ representation agrees qualitatively with the one obtained in the adjoint representation presented in the main text. The value of the spin gap differs, but not enormously. Unlike ${\rm Im} \cD(\bq,\omega)$, the imaginary part of $\Pi^s(\bq,\omega)$ is not a simple delta function, not even for $T \to 0$. There is rather a continuum of excitations above a certain momentum dependent threshold.
At finite temperatures there is also a low frequency continuum in ${\rm Im} \Pi^s(\bq,\omega)$, which is not present in ${\rm Im} \cD(\bq,\omega)$.



\bibliography{spin.bib}

\end{document}